\documentclass[12pt, reqno]{amsart}
\usepackage{graphicx,amsmath,amssymb,epsfig}
\usepackage{amsfonts}
\usepackage{geometry}
\usepackage{setspace}
\usepackage{lscape}
\usepackage{caption}
\usepackage{rotating}
\usepackage{xcolor}

\usepackage{amsfonts}
\theoremstyle{plain}

\newtheorem{theorem}{Theorem}

\newtheorem{algorithm}{Algorithm}

\newtheorem{condition}{Assumption}

\newtheorem{lemma}{Lemma}

\numberwithin{equation}{section}

\newcommand{\Gn}{\mathbb{G}_n}

\renewcommand{\Pr}{\mathbb{P}}
\newcommand{\Ep}{{\mathrm{E}}}
\newcommand{\En}{{\mathbb{E}_n}}
\newcommand{\vv}{\vartheta}
\newcommand{\V}{\mathcal{V}}
\newcommand{\D}{\mathcal{D}}
\newcommand{\W}{\mathcal{W}}
\newcommand{\Z}{\mathcal{Z}}
\newcommand{\R}{\mathcal{R}}
\newcommand{\T}{\mathcal{T}}

\begin{document}

\title[ ]{Quantile Regression with Censoring and Endogeneity}
\author[ ]{Victor Chernozhukov$^\dag$ \ \ Iv\'an
Fern\'andez-Val$^\S$ \ \ Amanda Kowalski$^\ddag$} \date{\today.  We
thank Denis Chetverikov and Sukjin Han for excellent comments and
capable research assistance. We are grateful to Richard Blundell for
providing us the data for the empirical application. We thank
the editor Cheng Hsiao, two referees, and seminar participants at EIEF, Georgetown, Rochester, and Penn State for useful comments.
We gratefully acknowledge research support from the NSF}

\thanks{\noindent $^\dag$ Department of Economics, MIT, 50 Memorial Drive, Cambridge, MA 02142,
vchern@mit.edu.}
\thanks{ $\S$ Boston University, Department of Economics, 270 Bay State Road,Boston, MA
02215, ivanf@bu.edu.}
\thanks{ $\ddag$ Department of Economics, Yale University, 37 Hillhouse Avenue, New Haven, CT
06520, and NBER, kowalski@nber.org.}

\maketitle

\begin{abstract} In this paper, we develop a new censored quantile
instrumental variable (CQIV) estimator and describe its properties
and computation. The CQIV estimator combines Powell (1986) censored
quantile regression (CQR) to deal  with censoring,
with a control variable approach to incorporate endogenous
regressors. The CQIV estimator is obtained in two stages that are
nonadditive in the unobservables. The first stage estimates a
nonadditive model with infinite dimensional parameters for the
control variable, such as a quantile or distribution regression
model. The second stage estimates a nonadditive censored quantile
regression model for the response variable of interest, including
the estimated control variable to deal with endogeneity. For
computation, we extend the algorithm for CQR developed by
Chernozhukov and Hong (2002) to incorporate the estimation of the
control variable. We give generic regularity conditions for
asymptotic normality of the CQIV estimator and for the validity of
resampling methods to approximate its asymptotic distribution. We
verify these conditions for quantile and distribution regression
estimation of the control variable. Our analysis covers two-stage (uncensored) quantile
regression with nonadditive first stage as an important special case. We illustrate the computation
and applicability of the CQIV estimator with a Monte-Carlo numerical
example and an empirical application on estimation of Engel curves for alcohol.

\end{abstract}

\newpage

\section{Introduction}

Censoring and endogeneity are common problems in data analysis. For
example, income survey data are often censored due to top-coding and many economic
variables such as hours worked, wages and expenditure shares are
naturally bounded from below by zero. Endogeneity is also an
ubiquitous phenomenon both in experimental studies due to partial
noncompliance (Angrist, Imbens, and Rubin, 1996), and in
observational studies due to simultaneity (Koopmans and Hood, 1953),
measurement error (Frish, 1934), sample selection (Heckman, 1979) or
more generally to  relevant omitted variables.
Censoring and endogeneity often come together in economic applications. For example,  both of them arise in the estimation of Engel curves for alcohol -- the relationship between the share of
expenditure on alcohol and the household's budget. For this
commodity, a significant fraction of households report zero expenditure,
and economic theory suggests that the total expenditure and its
composition are jointly determined in the consumption decision of
the household. Either
censoring or endogeneity lead to inconsistency of traditional mean
and quantile regression estimators by inducing correlation between
regressors and unobservables. We introduce a quantile regression
estimator that deals with both problems and name this estimator the
censored quantile instrumental variable (CQIV) estimator.

Our procedure deals with censoring semiparametrically through the
conditional quantile function following Powell (1986). \ This
approach avoids the strong parametric assumptions of traditional
Tobit  estimators. \ The key ingredient here is the equivariance
property of quantile functions to monotone transformations such as
censoring.  \ Powell's censored quantile regression estimator,
however, has proven to be difficult to compute. \ We address this
problem using the computationally attractive algorithm of
Chernozhukov and Hong (2002). \ An additional advantage of focusing
on the conditional quantile function is that we can capture heterogeneous effects  across
the distribution by computing CQIV at
different quantiles  (Koenker, 2005). The traditional Tobit
framework rules out this heterogeneity by imposing a location shift
model.

We deal with endogeneity using a control variable approach.  The
basic idea is to add a variable to the regression such that, once we
condition on this variable, regressors and unobservables become
independent.  This so-called control variable is usually
unobservable and needs to be estimated in a first stage.  Our main
contribution here is to allow for semiparametric models with
infinite dimensional parameters and nonadditive unobservables, such as
quantile regression and distribution regression, to model and
estimate the first stage and back out the control variable.  This
part of the analysis  constitutes the main theoretical difficulty
because the first stage estimators do not live in spaces with nice
entropic properties, unlike, for example, in Andrews (1994) or Newey
(1994). To overcome this problem, we develop a new technique to
derive asymptotic theory for two-stage procedures with plugged-in
first stage estimators that, while not living in Donsker spaces
themselves, can be suitably approximated by random functions that
live in Donsker spaces. This technique applies to semiparametric two-stage estimators where the
two stages can be nonadditive in the unobservables. CQIV is an example where the first
stage estimates a nonadditive quantile or distribution regression model for the control variable, whereas the second stage
estimates a nonadditive censored quantile regression model, including the estimated control
variable to deal with endogeneity. Two-stage (uncensored) quantile regression with distribution or quantile regression in the first stage is an important special case of CQIV.

We analyze the theoretical properties of the CQIV estimator in large
samples. \ Under suitable regularity conditions, CQIV is
$\sqrt{n}$-consistent and has a normal limiting distribution. \ We
characterize the expression of the asymptotic variance. \ Although
this expression can be estimated using standard methods, we find it
more convenient to use resampling methods for inference.\ We focus
on weighted bootstrap because the proof of its consistency is not overly
complex following the strategy set forth by Ma and Kosorok (2005). \ We
give regularity conditions for the consistency of weighted bootstrap
to approximate the distribution of the CQIV estimator. \ For our
leading cases of quantile and distribution regression estimation of
the control variable, we provide more primitive assumptions that
verify the regularity conditions for asymptotic normality and
weighted bootstrap consistency.  \ The verification of these
conditions for two-stage censored and uncensored quantile regression estimators based on quantile and distribution regression estimators of
the first stage is new to the best of our knowledge.

The CQIV estimator is simple to compute using standard statistical
software.\footnote{We have developed a Stata command to implement
the methods developed in this paper (see Chernozhukov,
Fernandez-Val, Han, and Kowalski 2011).  It is  available at
http://EconPapers.repec.org/RePEc:boc:bocode:s457478.} We
demonstrate its implementation through Monte-Carlo simulations and an empirical application to the estimation of Engel
curves for alcohol. The
results of the Monte-Carlo exercise demonstrate that the performance
of CQIV is comparable to that of Tobit IV in data generated to
satisfy the Tobit IV assumptions, and it outperforms Tobit IV in data that do not satisfy these assumptions. The results of the application to Engel curves demonstrate the importance of accounting
for endogeneity and censoring in real data. Another application of our CQIV estimator to the
estimation of the price elasticity of expenditure on medical care
appears in Kowalski (2009).

\subsection{Literature review.} There is an extensive previous
literature on the control variable approach to deal with endogeneity
in models without censoring. Hausman (1978) and Wooldridge (2010)
discussed parametric triangular linear and nonlinear models. \
Newey, Powell, and Vella (1999) described the use of this approach
in nonparametric triangular systems of equations for the conditional
mean, but limited the analysis to models with additive unobservables both
in the first and the second stage. \ Blundell and Powell (2004) and Rothe (2009) applied the
control variable approach to semiparametric binary response models. \ Lee (2007) set forth an
estimation strategy using a control variable approach for a
triangular system of equations for conditional quantiles with an
additive nonparametric first stage.  \ Imbens and Newey (2002, 2009)
extended the analysis to triangular nonseparable models with
nonadditive unobservables in both the first and second stage. They
focused on identification and nonparametric estimation rates for
average, quantile and policy effects. Our paper complements Imbens
and Newey (2002, 2009) by providing inference methods and allowing
for censoring. \ Chesher (2003) and Jun (2009) considered local
identification and semiparametric estimation of uncensored
triangular quantile regression models with a nonseparable control
variable. \  Relative to CQIV, these methods have 
the advantage that they impose less structure in the model at the cost of
slower rates of convergence in estimation. In particular, they leave the dependence on the control variable unspecified, whereas CQIV uses a  flexible parametric specification.  \ While the previous papers focused on triangular
models, Blundell and Matzkin (2010) have recently derived conditions
for the existence of control variables in nonseparable simultaneous
equations models. \ We refer also to Blundell and Powell (2003) and Matzkin (2007) for excellent
comprehensive reviews of results on semi  and nonparametric identification and estimation of
triangular and simultaneous equations models.

Our work is also closely related to Ma and Koenker (2006). \ They
considered identification and estimation of quantile effects without
censoring using a parametric control variable. \ Their parametric
assumptions rule out the use of nonadditive models with infinite
dimensional parameters in the first stage, such as quantile and
distribution regression models. \ In contrast,
our approach is specifically designed to handle the latter, and in
doing so, it puts the first stage and second stage models on
equally flexible footing.   \  Allowing for a nonadditive infinite
dimensional control variable makes the analysis of the asymptotic
properties of our estimator very delicate and requires developing
new proof techniques because of the difficulties discussed above.



For models with censoring and exogenous regressors, Powell (1986),
Fitzenberger (1997),  Buchinsky and Hahn (1998), Khan and Powell
(2001), Chernozhukov and Hong (2002), Honor\'e, Khan and Powell
(2003), and Portnoy (2003) developed quantile regression methods. \
The literature on models combining both endogeneity and censoring is
more sparse. \ Smith and Blundell (1986) pioneered the use of the
control variable approach to estimate a triangular parametric
additive location model. \ More recently, Blundell
and Powell (2007) proposed an alternative censored quantile
instrumental variable estimator building on Chen and Khan (2001).
Compared to our estimator, Blundell and Powell estimator assumes
additive unobservables in the first and second stages, but permits a flexible local
nonparametric endogeneity correction in the second stage.  \ Hong and Tamer (2003)  and Khan and
Tamer (2006) also considered censored regression models with endogenous regressors. They dealt
with endogeneity with an instrumental variable quantile approach that is not nested with the
control variable approach used here; see Blundell and Powell (2003) for a comparison  of these two approaches. They dealt with censoring using a more flexible moment inequality formulation that allows for  endogenous censoring and partial identification, but that  leads to a more complicated estimator.   A referee has pointed to us the possibility of applying the control variable approach as pursued in this paper to Buchinsky and Hahn (1998) estimator to deal with endogenous regressors.
 We believe that this is indeed possible using the results of this paper, though we leave formal developments to future work. 
 

Relative to the previous literature, the paper makes three main contributions.  First, it develops
a two-stage quantile regression estimator for a triangular  nonseparable model where the first stage is nonadditive in the unobservables. Our analysis here builds on Chernozhukov, Fernandez-Val, and Galichon (2010) and Chernozhukov, Fernandez-Val, and Melly (2013), which established the properties of the quantile and distribution regression estimators that we use in the first stage. The theory for the second stage estimator, however, does not follow from these results using standard techniques due to the dimensionality and entropy properties of the first stage estimators. Second, it adapts the two-stage quantile regression estimator to models with censoring by extending Chenozhukov and Hong (2002) algorithm to the presence of  a generated regressor (control variable). Third, it establishes the validity of weighted bootstrap for two-stage censored and uncensored quantile regression estimators where the first stage is estimated by quantile or distribution regression.

\subsection{Plan of the paper.} The rest of the paper is organized
as follows. In Section \ref{model}, we present the CQIV model and
develop estimation and inference methods for the parameters of
interest of this model. In Sections \ref{computation} and
\ref{montecarlo}, we describe the associated computational
algorithms and present results from a Monte-Carlo simulation
exercise. In Section \ref%
{engel}, we present an empirical application of CQIV to Engel
curves. In Section \ref{conclusion}, we provide conclusions and
discuss potential empirical applications of CQIV. The proofs of the
main results are given in the appendix.

\vfill 

\section{Censored Quantile Instrumental Variable Regression}

\label{model}

\subsection{The Model}

We consider the following triangular system of quantile equations:
\begin{eqnarray}
Y &=& \max (Y^{\ast },C),  \label{2.1} \\
Y^{\ast } &=& Q_{Y^{\ast }}(U \mid D,W,V), \label{2.2} \\
D &=& Q_{D}(V \mid W,Z). \label{2.3}
\end{eqnarray}%
In this system, $Y^{\ast }$ is a continuous latent response
variable, the observed variable $Y$ is obtained by censoring
$Y^{\ast }$ from below at the level determined by the variable $C$,
$D$ is the continuous regressor of interest, $W$ is a vector of
covariates, possibly containing $C$, $V$ is a latent unobserved
regressor that accounts for the possible endogeneity of $D$, and $Z$
is a vector of ``instrumental variables'' excluded from
(\ref{2.2}).\footnote{We focus on left censored response variables
without loss of generality. If $Y$ is right censored at $C$, $Y =
\min(Y^{\ast },C)$, the analysis of the paper applies without change
to $\widetilde Y = - Y$, $\widetilde Y^{\ast} = - Y^{\ast}$,
$\widetilde C = - C$, and $Q_{\widetilde Y^{\ast }} = - Q_{Y^{\ast
}}$, because $\widetilde Y = \max(\widetilde Y^{\ast }, \widetilde
C)$.}  \ The uncensored case is covered by making $C$ arbitrarily small.

The function $u\mapsto Q_{Y^{\ast }}(u \mid D,W,V)
$ is the conditional quantile function of $Y^{\ast }$ given $(D,W,V)$; and $%
v\mapsto Q_{D}(v \mid W,Z)$ is the conditional quantile function of
the regressor $D$ given $(W,Z)$. $\ $Here, $U$ is a Skorohod
disturbance for $Y$ that satisfies the independence assumption
\begin{equation*}
U\sim U(0,1) \mid D,W,Z,V,C,
\end{equation*}%
and $V$ is a Skorohod disturbance for $D$  that satisfies
\begin{equation*}
V\sim U(0,1)  \mid W,Z,C.
\end{equation*}%
In the last two equations, we make the assumption that the censoring
variable $C$ is independent of the disturbances $U$ and $V$. This
variable can, in principle, be included in $W$.  \
To recover the conditional quantile function of the latent response
variable in equation (\ref{2.2}), it is important to condition on an
unobserved regressor $V$ which plays the role of a \textquotedblleft
control variable.\textquotedblright \ Equation (\ref{2.3}) allows us
to recover this unobserved regressor as a residual that explains
movements in the variable $D$, conditional on the set of instruments
and other covariates. \  The main identification conditions are the exclusion restriction of $Z$ in equation (\ref{2.2}), and the
relevance condition of $Z$ in equation (\ref{2.3}). These conditions permit  $V$ to have independent variation of
$D$ and $W$. \


 An example of a structural model that has the triangular representation
(\ref{2.2})-(\ref{2.3}) is the  system of equations
\begin{eqnarray}
Y^{\ast } &=&   \beta_1 D + \beta_2'W + (\beta_3 D + \beta_4'W) \epsilon, \label{eq: normal1} \\
D &=&  \pi_1'Z + \pi_2'W + (\pi_3'Z + \pi_4'W)\eta, \label{eq: normal2}
\end{eqnarray}%
where $(\epsilon,\eta)$ are jointly standard bivariate normal with correlation $\rho'W$ conditional on $(W,Z,C)$,  $ (\beta_3 D + \beta_4'W) > 0$ a.s., and $(\pi_3'Z + \pi_4'W) > 0$ a.s.  By the properties of the normal distribution, $\eta = \Phi^{-1}(V)$ with $V \sim U(0,1)$ independent of $(W,Z,C),$ and $\epsilon = (\rho'W) \Phi^{-1}(V) + [1 - (\rho'W)^2]^{1/2} \Phi^{-1}(U)$ with $U \sim U(0,1)$ independent of $(W,Z,C,V,D),$ where $\Phi^{-1}$ is the inverse distribution function of the standard normal.
The corresponding conditional
quantile functions have the form of (\ref{2.2}) and (\ref{2.3}) with
\begin{eqnarray*}
Q_{Y^{\ast }}(U\mid D,W,V) &=& \beta_1 D + \beta_2'W + (\beta_3 D + \beta_4'W)\{  (\rho'W) \Phi^{-1}(V) + [1 - (\rho'W)^2]^{1/2} \Phi^{-1}(U)\} ,  \\
Q_{D}(V\mid W,Z) &=&  \pi_1'Z + \pi_2'W + (\pi_3'Z + \pi_4'W)\Phi^{-1}(V). \end{eqnarray*} Both of these quantile functions are nonadditive in $U$ and $V$. We use a simplified
version of the system (\ref{eq: normal1})--(\ref{eq: normal2}) to generate the data for the numerical examples in
Section \ref{montecarlo}.
%

In the system  (\ref{2.1})--(\ref{2.3}), the  observed
response variable has the quantile representation
\begin{equation}
Y=Q_{Y}(U\mid D,W,V,C)=\max (Q_{Y^{\ast }}(U\mid D,W,V),C),
\label{2.5}
\end{equation}%
by the equivariance property of the quantiles to monotone
transformations.
Whether the response of interest is the latent or observed variable
depends on the source of censoring (e.g., Wooldridge, 2010, Chap. 17). When
censoring is due to data limitations such as top-coding, we are
often interested in the conditional quantile function of the latent
response variable $Q_{Y^{\ast}}$ and marginal effects derived from
this function. For example, in the system (\ref{eq:
normal1})--(\ref{eq: normal2}) the marginal effect of the endogenous
regressor $D$ evaluated at $(D,W,V,U) = (d,w,v,u)$ is
$$\partial_d Q_{Y^{\ast}}(u \mid d,w,v) =
\beta_1 + \beta_3 \{  (\rho'w) \Phi^{-1}(v) + [1 - (\rho'w)^2]^{1/2} \Phi^{-1}(u)\},$$ which corresponds
to the ceteris paribus effect of a marginal change of $D$ on the
latent response $Y^{\ast}$ for individuals with $(D,W,V,U) =
(d,w,v,u)$. When the censoring is due to economic or
behavioral reasons such are corner solutions, we are often
interested in the conditional quantile function of the observed
response variable $Q_{Y}$ and marginal effects derived from this
function. For example,  the marginal effect of the endogenous regressor $D$
evaluated at $(D,W,V,U,C) = (d,w,v,u,c)$ is
$$\partial_d Q_{Y}(u \mid d,w,v,c) =
1\{Q_{Y^{\ast}}(u \mid d,w,v) > c\}\partial_d Q_{Y^{\ast}}(u \mid d,w,v),$$ which corresponds to the ceteris paribus
effect of a marginal change of $D$ on the observed response $Y$ for
individuals with $(D,W,V,C,U) = (d,w,v,c,u)$.
Since either of the marginal effects might depend  on individual
characteristics, average marginal effects or marginal effects
evaluated at interesting values are often reported.

\subsection{Generic Estimation}
\label{subsec: estimation}

To make estimation both practical and realistic, we impose a
flexible semiparametric restriction on the functional form of the
conditional quantile function in (\ref{2.2}). In particular, we
assume that
\begin{equation}\label{define 1 model}
Q_{Y^{\ast} }(u\mid D,W,V)=X^{\prime }\beta_0 (u),\ \ X = x(D,W,V),
\end{equation}%
where $x(D,W,V)$ is a  vector of transformations of the initial
regressors $(D,W,V)$. \ The transformations could be, for example,
polynomial, trigonometric, B-spline or other basis functions that
have good approximating properties for economic problems.
For the control variable, it is convenient to take a  strictly monotonic transformation to adjust the location and scale (Newey, 2009), and
to include interactions with the basis of $D$ and $W$ to account for nonseparabilities.\footnote{For example,  the
transformation $\Phi^{-1}(V),$ where $\Phi$ is the distribution function of the standard normal, yields the control variable in the  system (\ref{eq: normal1})--(\ref{eq: normal2}).}
An important property of this functional form is linearity in
parameters, which is very convenient for computation.
The resulting conditional quantile function of the censored random
variable
$$
Y= \max(Y^*,C),
$$
 is given by
\begin{equation}\label{define 2 model}
Q_{Y}(u\mid D,W,V,C)=\max (X^{\prime }\beta_0 (u),C).
\end{equation}%
This is the standard functional form for the censored quantile
regression (CQR) first derived by Powell (1984) in the exogenous
case.


Given a random sample $\{Y_i,D_i,W_i,Z_i,C_i\}_{i = 1}^{n}$, we form
the estimator for the parameter $\beta_0(u)$ as
\begin{equation}\label{eq: cqiv}
\widehat{\beta }(u)=\arg \min_{\beta \in \mathbb{R}^{\dim(X)}}\frac{1}{n}\sum_{i=1}^{n} 1(%
\widehat{S}_{i}^{\prime }\widehat{\gamma}(u)\geq\varsigma(u)) T_i  \rho
_{u}(Y_{i}-\widehat X_{i}^{\prime }\beta),
\end{equation}%
where $\rho _{u}(z)=(u-1(z<0))z$ is the asymmetric absolute loss
function of
Koenker and Bassett (1978), $\widehat{X}_{i}=x(D_{i},W_{i},\widehat{V_{i}})$%
, $\widehat{S}_{i}=s(\widehat{X}_{i},C_i),$ $s(X,C)$ is a vector
of transformations of $(X,C)$, $\varsigma(u)$ is a positive cut-off,  $\widehat{V_{i}}$ is an estimator of $%
V_{i}$, and $ T_i $ is an exogenous trimming indicator defined in Assumption \ref{condition: trimming} that we include for technical reasons.  \  The estimator in (\ref{eq: cqiv}) adapts the algorithm for the CQR estimator
developed in Chernozhukov and Hong (2002) to deal with endogeneity. \ This algorithm is
based on the property of the model
$$
P(Y \leq X'\beta_0(u) \mid X, C, X'\beta_0(u) > C) = P(Y^* \leq X'\beta_0(u) \mid X,C,X'\beta_0(u) > C) = u,
$$
provided that $P(X'\beta_0(u) > C) > 0.$ In other words, $X'\beta_0(u)$ is the conditional $u$-quantile of the observed outcome for the observations for which $X'\beta_0(u) > C,$ i.e., the conditional $u$-quantile of the latent outcome is above the censoring point. These observations change with the quantile index $u$ and may include censored observations. We refer to them as the ``$u$-quantile uncensored'' observations.
The multiplier $1(\widehat{S}_{i}^{\prime
}\widehat{\gamma}(u)\geq\varsigma(u))$ is a selector that predicts if observation $i$ is $u$-quantile uncensored. We formally state the conditions on this selector in Assumption \ref{condition: selector}. The estimator in $(\ref{eq: cqiv})$ may also be seen as
a computationally attractive approximation to Powell estimator
applied to our case:
\begin{equation*}
\widehat{\beta }_{p}(u)=\arg \min_{\beta \in \mathbb{R}^{\dim(X)}}\frac{1}{n}%
\sum_{i=1}^{n}  T_i \rho _{u}[Y_{i}-\max (\widehat{X}_{i}^{\prime}\beta ,
C_i)].
\end{equation*}

The CQIV estimator will be computed using an iterative procedure
where each step will take the form specified in equation (\ref{eq:
cqiv}). We start selecting the set of $u$-quantile uncensored
observations by estimating the
conditional probabilities of censoring using a flexible binary
choice model. These observations have conditional probability of
censoring lower than the quantile index $u$ because of the equivalence of the events $\{X'\beta_0(u) > C\} \equiv \{P(Y^* \leq C \mid X,C) < u\}$. We estimate the linear
part of the conditional quantile function, $X_i'\beta_0(u)$, on the
sample of $u$-quantile uncensored observations by standard quantile
regression. Then, we update the set of $u$-quantile uncensored
observations by selecting those observations with conditional
quantile estimates that are above their censoring points, $X_i'\widehat\beta(u) > C_i$,  and
iterate. We provide more practical implementation details in the
next section.

The control variable $V$ can be estimated in several ways. Note that
if $Q_D(v \mid W,Z)$ is invertible in $v$, the control variable has
two equivalent representations:
\begin{equation}\label{define 3 model}
V= \vartheta_0(D,W,Z)\equiv F_{D}(D \mid W,Z) \equiv
Q_{D}^{-1}(D\mid W,Z).
\end{equation}%
For any estimator of $F_{D}(D \mid W,Z)$ or  $Q_{D}^{-1}(V\mid W,Z)$,
denoted by $\widehat F_{D}(D \mid W,Z)$ or  $\widehat {Q_{D}^{-1}}(V\mid
W,Z)$, based on any parametric or semiparametric functional form,
the resulting estimator for the control variable is
\begin{equation*}
\widehat{V}=\widehat{\vartheta}(D,W,Z)\equiv \widehat{F}_{D}(D\mid
W,Z) \text{ or } \widehat{V}=\widehat{\vartheta}(D,W,Z)\equiv
\widehat{Q^{-1}_{D}}(D\mid W,Z).
\end{equation*}%

Here we consider several examples: in the classical additive
location model,  $Q_{D}(v\mid W,Z)=R^{\prime }\pi_0
+Q_{V}(v),$ where $Q_{V}$ is a quantile function, and $R = r(W,Z)$
is a vector collecting transformations of $W$ and $Z$. The control
variable is
\begin{equation*}
V=Q_{V}^{-1}(D-R^{\prime }\pi_0 ),
\end{equation*}%
which can be estimated by the empirical CDF of the least squares
residuals. Chernozhukov, Fernandez-Val and Melly (2013) developed
asymptotic theory for this estimator. If $D \mid W,Z \sim N(R'\pi_0,
\sigma^2)$, the control variable has the  parametric form $V =
\Phi^{-1}([D-R^{\prime }\pi_0]/\sigma)$, where $\Phi$ denotes
the distribution function of the standard normal distribution. This
control variable can be estimated by plugging in estimates of the
regression coefficients and  residual variance.

In a nonadditive quantile regression model, we have that
$Q_{D}(v\mid W,Z)=R^{\prime }\pi_0 (v),$ and
\begin{equation*}
V= Q_{D}^{-1}(D\mid W,Z) = \int_{(0,1)} 1\{R^{\prime }\pi_0 (v)\leq
D\}dv.
\end{equation*}%
The estimator takes the form
\begin{equation}
\widehat{V}= \tau + \int_{(\tau,1-\tau)}1\{R^{\prime }\widehat{\pi }(v)\leq D\}dv,
\label{equation: fe_qr_estimator}
\end{equation}%
where $\widehat{\pi }(v)$ is the Koenker and Bassett (1978) quantile
regression estimator, $\tau$ is small positive trimming cut-off that avoids estimation of tail quantiles (Koenker,
2005, p. 148),  and the integral can be approximated
numerically using a finite grid of quantiles. The use of the
integral representation of $Q_{D}^{-1}$ with respect to $Q_{D}$ is convenient to avoid potential noninvertibility of $\widehat{Q}_D$ caused by nonmonotonicity of $v \mapsto R^{\prime }\widehat{\pi }(v)$.
Chernozhukov, Fernandez-Val, and Galichon (2010) developed asymptotic theory for
this estimator.

We can also estimate $\vartheta_0$ using distribution regression. In
this case we consider a semiparametric model for the conditional
distribution of $D$ to construct a control variable
$$
V  = F_D(D \mid W,Z) = \Lambda(R' \pi_0 (D)),
$$
where $\Lambda$ is a probit or logit link function. The estimator
takes the form
\begin{equation}\label{equation: fe_dr_estimator}
\widehat V = \Lambda(R' \widehat{\pi}(D)),
\end{equation}
where $\widehat{\pi}(d)$ is the maximum likelihood estimator of
$\pi_0(d)$ at each $d$ (see, e.g., Foresi and Peracchi, 1995, and
Chernozhukov, Fernandez-Val and Melly, 2013). Chernozhukov,
Fernandez-Val and Melly (2013) developed asymptotic theory for this
estimator.

The classical additive location model is an special case of the quantile regression model, where only the coefficient of the intercept varies across quantiles. The quantile and distribution regression models are flexible in the sense that by allowing for a sufficiently rich $R$, we can approximate any conditional distributions arbitrarily well.  These models are not nested, so they cannot be ranked on the basis of generality. We refer to Chernozhukov, Fernandez-Val and Melly (2013) for a detailed comparison of these models.

\subsection{Regularity Conditions for Estimation}

In what follows, we shall use the following notation.  We let the
random vector $A= (Y,D,W,Z,C,X,V)$ live on some probability space
$(\Omega_0, \mathcal{F}_0, P)$. Thus, the probability measure $P$
determines the law of $A$ or any of its elements. We also let
$A_1,...,A_n$, i.i.d. copies of $A$, live on the complete
probability space $(\Omega, \mathcal{F}, \Pr)$, which contains the
infinite product of $(\Omega_0, \mathcal{F}_0, P)$. Moreover, this
probability space can be suitably enriched to carry also the random
weights that will appear in the weighted bootstrap. The distinction
between the two laws $P$ and $\Pr$ is helpful to simplify the
notation in the proofs and in the analysis. Calligraphic letters such
as $\mathcal{Y}$ and $\mathcal{X}$ denote the closures of the supports of $Y$ and
$X$; and $\mathcal{YX}$ denotes the closure of the joint support of $(Y,X)$.
Unless
explicitly mentioned, all functions appearing in the statements are
assumed to be measurable.

We now state formally the assumptions. The first assumption is our
model.

\begin{condition}[Model]  We observe $\{Y_{i},D_{i},W_{i},Z_{i},C_{i}\}_{i =
1}^{n}$, a sample of size $n$ of independent and identically
distributed observations from the random vector $(Y,D,W,Z,C),$ which
obeys the model assumptions
\begin{eqnarray*}
&& Q_{Y}(u \mid D,W,Z,V,C) = Q_{Y}(u \mid X,C) = \max(X'\beta_0(u),C), \ \ X = x(D,W,V), \\
&& V = \vartheta_0(D,W,Z) \equiv F_{D}(D \mid W,Z) \sim U(0,1) \mid
W,Z.
\end{eqnarray*}
\end{condition}

We define a trimming indicator that equals one whenever $D$ lies in a region that exclude extreme values.
The purpose of the trimming is to avoid the far tails in the modeling and estimation of the control variable in the first stage.   We consider a fixed trimming rule, which greatly simplifies the derivation of the asymptotic properties. Alternative random, data driven rules are also possible at the cost of more complicated proofs. We discuss the choice of the trimming rule in Section \ref{computation}.
\begin{condition}[Trimming indicator]\label{condition: trimming}
The tail trimming indicator has the form
\begin{equation*}
T = 1(D \in \overline{\D}),
\end{equation*}%
where $\overline{\D} = [\underline{d}, \overline{d}]$ for some $-\infty < \underline{d} < \overline{d} < \infty$, such that $P(T = 1)>0$.
\end{condition}

Throughout the paper we use bars to denote trimmed supports with respect to $D$, e.g.,  $\overline{\D\W\Z} = \{(d,w,z) \in \D\W\Z :  d \in \overline{\D} \}$,
and $\overline{\V} = \{\vv_0(d,w,z): (d,w,z) \in \overline{\D\W\Z}\}$. The next assumption imposes compactness and smoothness conditions. Compactness is imposed over the trimmed supports and can be relaxed at the cost of more complicated and cumbersome proofs.  Moreover, we do not require compactness of the support of $Y$, which is important to cover the tobit model. The smoothness conditions are fairly tight.

\begin{condition}[Compactness and smoothness] \label{condition:model} (a) The set
$\overline{\mathcal{DWZCX}}$  is compact. (b) The endogenous regressor $D$ has
a continuous conditional density $f_{D}(\cdot \mid w,z)$ that is
 bounded above by a constant uniformly in $(w,z) \in \overline{\mathcal{WZ}}$. (c) The random variable $Y$ has a
conditional density $f_{Y}(y \mid x,c)$ on $(c,\infty)$ that is
uniformly continuous in $y \in (c,\infty)$ uniformly in $(x,c) \in
\overline{\mathcal{XC}}$, and bounded above by a constant uniformly in $(x,c)
\in \overline{\mathcal{XC}}$. (d) The derivative vector $\partial_v x(d,w,v)$
exists and its components are uniformly continuous in $v\in\overline{\V}$
uniformly in $(d,w) \in \overline{\mathcal{DW}}$, and  are bounded in absolute
value by a constant, uniformly in $(d,w,v) \in \overline{\mathcal{DWV}}$.
\end{condition}

The following assumption is a high-level condition on the
function-valued estimator of the control variable. We assume that it
has a uniform asymptotic functional linear representation. The trimming device facilitates this assumption because it limits the convergence to a region that excludes the tails of the control variable.  Moreover, the function-valued estimator, while not necessarily living in a Donsker
class, can be approximated by a random function that does live in a
Donsker class. We will fully verify this condition for the case of
quantile regression and distribution regression under more primitive
conditions.  Let $T(d) := 1(d \in \overline{\D})$ and $\|f\|_{T,\infty} := \sup_{a \in \mathcal{A}} |T(d) f(a)|$ for any function $f : \mathcal{A} \mapsto \mathbb{R}$.

\begin{condition}[Estimator of the control variable]\label{condition:control}
We have an estimator of the control variable of the form
$\widehat{V}=\widehat{\vartheta}(D,W, Z)$  such that uniformly over
$\overline{\D\W\Z}$, (a)
\begin{equation*}
 \sqrt{n}(\widehat{\vartheta}(d,w,z)-\vartheta_0(d,w,z)) = \frac{1}{\sqrt{n}}%
\sum_{i=1}^{n} \ell(A_i, d,w,z) + o_{\Pr}(1), \ \ \Ep_{P}[\ell(A,
d,w,z)]=0,
\end{equation*}%
where $\Ep_{P}[T \ell(A, D,W,Z)^2] < \infty$ and $\| \frac{1}{\sqrt{n}}%
\sum_{i=1}^{n} \ell(A_i, \cdot)\|_{T,\infty} = O_{\Pr}(1)$, and (b)
$$
\|\widehat  \vartheta -  \widetilde \vartheta\|_{T,\infty} = o_{\Pr}(1/\sqrt{n}), \ \ \text{ for } \ \  \widetilde \vartheta \in \Upsilon
$$
with probability approaching one, where the covering entropy of the function class $\Upsilon$ is not too high, namely
$$
\log N (\epsilon, \Upsilon, \|\cdot\|_{T,\infty}) \lesssim  1/(\epsilon \log^4 (1/
\epsilon)), \text{ for  all } 0 < \epsilon < 1.
$$
\end{condition}

The following assumptions are on the $u$-quantile uncensored selector. The first part is a
high-level condition on the estimator of the selector.  The second
part is a smoothness condition on the index that defines the
selector. We shall verify that the CQIV estimator can act as a
legitimate selector itself. Although the statement is involved, this
condition can be easily satisfied as explained below.

\begin{condition}[Quantile-uncensored selector]\label{condition: selector}
(a) The selection rule has the form
\begin{equation*}
1[s(x(D,W,\widehat V),C)^{\prime }\widehat{\gamma}(u)\geq\varsigma(u)],
\end{equation*}%
for some $\varsigma(u) > 0$, where $\widehat{\gamma}(u)\rightarrow
_{\Pr}\gamma_0(u)$ and, for some $\epsilon'>0$,
\begin{equation*}
1[S^{\prime }\gamma_0(u) \geq\varsigma(u)/2] T \leq  1[X'\beta_0(u)\geq C + \epsilon'] T \\
\leq  1[X'\beta_0(u)>C ] T \text{ $P$-a.e.,}
\end{equation*}%
where $S = s(X,C)$.
 (b) The set $\overline{\mathcal{S}}$  is compact. (c) The density
of the random variable $s(x(D,W,\vartheta(D,W,Z)),C)'\gamma$ exists
and is bounded above  by a constant, uniformly in $\gamma \in
\Gamma(u)$
 and in $\vartheta \in \Upsilon$, where $\Gamma(u)$ is an open neighborhood
 of $\gamma_0(u)$ and $\Upsilon$ is defined in Assumption \ref{condition:control}. (d) The
 components of the derivative vector $\partial_v s(x(d,w,v),c)$ are uniformly continuous at each
$v\in \overline{\V}$ uniformly in $(d,w,c) \in \overline{\mathcal{DWC}}$, and are
bounded in absolute value by a constant, uniformly in $(d,w,v,c) \in
\overline{\mathcal{DWVC}}$.
\end{condition}

The next assumption is a sufficient condition to guarantee local
identification of the parameter of interest as well as
$\sqrt{n}$-consistency and asymptotic normality of the estimator.

\begin{condition}[Identification and nondegeneracy]\label{condition: identification}
(a) The matrix
\begin{equation*}
J(u) := \Ep_P [f_{Y}(X^{\prime}\beta_0(u) \mid X,C) X X^{\prime} \
1( S' \gamma_0(u) \geq\varsigma(u) ) T]
\end{equation*}
is of full rank. (b) The matrix
\begin{equation*}
\Lambda (u) := \mathrm{Var}_P[ f(A) + g(A) \ ],
\end{equation*}
is finite and is of full rank, where
$$
f(A) := \{1(Y< X^{\prime}\beta_0(u)) - u\} X 1 (S^{\prime}\gamma_0(u)
\geq \varsigma(u)) T,
$$
and, for $\dot{X} = \partial_{v} x(D, W, v)|_{v= V}$,
$$
g(A) := \Ep_P[ f_{Y} (X'\beta_0(u) \mid X, C) X \dot X'\beta_0(u) 1
( S'\gamma_0(u) \geq \varsigma(u) ) T \ell(a,D,W,Z) ] \big|_{a=A}.
$$
\end{condition}



Assumption \ref{condition: selector}(a) requires the selector to find a  subset of the
$u$-quantile-censored observations, whereas Assumption \ref{condition: identification} requires the selector to find a nonempty subset.  Let  $\widehat \beta^0(u)$ be an
initial consistent estimator of $\beta_0(u)$ that uses
a selector based on a flexible model for the conditional probability of
censoring as described in Section  \ref{computation}. This model does not need to be correctly specified under
a mild separating hyperplane condition for the $u$-quantile uncensored
observations (Chernozhukov and Hong, 2002). Then, we update the
selector to $1[s(x(D,W,\widehat V),C)'\widehat \gamma(u) \geq \varsigma(u)],$
where $s(x(D,W,\widehat V),C) = [x(D,W,\widehat V)', C]'$, and $\widehat
\gamma(u) = [\widehat \beta^0(u)', -1]'$. The parameter $\varsigma(u)$ is a small
fixed cut-off that  ensures that the selector is asymptotically
conservative but nontrivial.  We provide guidelines for the choice of $\varsigma(u)$
in  Section \ref{computation} and show that the CQIV estimates are not very sensitive to this choice in the numerical examples of Section \ref{montecarlo}.



The full rank conditions of Assumption \ref{condition: identification}  hold if there are not perfectly collinear components in the vector $X = x(D,W, \vartheta_0(D,W,Z))$ and $P(S^{\prime}\gamma_0(u)
\geq \varsigma(u), T = 1) > 0$. To avoid reliance on functional form assumptions for $x$ and $\vartheta_0$, the noncollinearity  requires the exclusion restriction for $Z$  in Assumption 1, $Q_Y(u \mid D,W,Z,V,C) = Q_Y(u \mid D,W,V,C)$  a.s., and a global relevance or rank condition for $Z$ such as  $\text{Var}_P[\vartheta_0(D,W,Z) | D,W] > 0$ a.s.
Chesher (2003) and Jun (2009) impose local versions of the exclusion and relevance conditions for $Z$  at a point of interest for $V$. 

\subsection{Main Estimation Results}

The following result states that the CQIV estimator is consistent,
converges to the true parameter at a $\sqrt{n}$-rate, and is
normally distributed in large samples.

\begin{theorem}[Asymptotic distribution of CQIV]
Under the stated assumptions
\begin{equation*}
\sqrt{n}(\widehat\beta(u) - \beta_0(u)) \to_d N(0, J^{-1}(u) \Lambda
(u) J^{-1}(u)).
\end{equation*}
\end{theorem}
We can estimate the variance-covariance matrix $J^{-1}(u) \Lambda
(u) J^{-1}(u)$ using standard
methods and carry out analytical inference based on the normal
distribution. Estimators for the components of the variance can be
formed following Powell (1991) and Koenker (2005). However,
this is not very convenient for practice due to the complicated form
of these components and the need to estimate conditional densities.
Instead, we suggest using weighted bootstrap (Ma and Kosorok, 2005, Chen and Pouzo, 2009) and prove
its validity in what follows.

We focus on weighted bootstrap because the proof of its consistency is  not overly
complex, following the strategy set forth by Ma and Kosorok (2005). This bootstrap also has practical advantages
over nonparametric bootstrap to deal with discrete regressors with
small cell sizes, because it avoids having singular designs under the bootstrap data generating process.
Moreover, a particular version of the weighted bootstrap, with
exponentials acting as weights,  has a nice Bayesian interpretation
(Hahn, 1997, Chamberlain and Imbens, 2003).

To describe the weighted bootstrap procedure in our setting, we
first introduce the ``weights''.

\begin{condition}[Bootstrap weights] \label{condition:weights}
The weights
$(e_1, ..., e_n)$ are i.i.d. draws from a random variable $e \geq
0$, with $\Ep_P[e] = 1$, $\mathrm{Var}_P[e] = 1,$ and $\Ep_P |e|^{2+\delta} < \infty$ for some $\delta > 0$; live on the
probability space $(\Omega, \mathcal{F},\Pr)$; and are independent of
the data $\{Y_{i},D_{i},W_{i},Z_{i},C_i\}_{i=1}^n$ for all $n$.
\end{condition}

\noindent \textbf{Remark 1} (Bootstrap weights). The chief and
recommended example of bootstrap weights is given by $e$ set to be
the standard exponential random variable. Note that for other
positive random variables with $\Ep_P[e] = 1$ but $\mathrm{Var}_P[e]
> 1$, we can take the transformation $\tilde e = 1 + (e -
1)/\mathrm{Var}_P[e]^{1/2}$, which satisfies $\tilde e \geq 0$,
$\Ep_P[\tilde e] =1$, and $\mathrm{Var}_P[\tilde e]=1$.

\medskip

The weights act as sampling weights in the bootstrap procedure. In
each repetition, we draw a new set of weights $(e_1,\ldots, e_n)$
and recompute the CQIV estimator in the weighted sample. We refer to
the next section for practical details, and here we define the
quantities needed to verify the validity of this bootstrap scheme.
Specifically, let $\widehat V_{i}^e$ denote the estimator of the
control variable for observation $i$ in the weighted sample, such as
the quantile regression or distribution regression based estimators
described below. The CQIV estimator in the weighted
sample solves
\begin{equation}\label{eq: weighted cqiv}
\widehat{\beta }^{e}(u) =\arg \min_{\beta \in \mathbb{R}^{\dim(X)}}\frac{1}{n}\sum_{i=1}^{n} e_i 1(%
\widehat{\gamma}(u)^{\prime}\widehat{S}_{i}^{e} \geq\varsigma(u)) T_i \rho
_{u}(Y_{i}-\beta^{\prime}\widehat X_{i}^{e }),
\end{equation}
where $\widehat{X}_{ i}^e = x(D_i, W_i, \widehat V_{ i}^e)$,
$\widehat{S}_{ i}^e = s(\widehat{X}_{ i}^e, C_i)$,
 and
$\widehat{\gamma}(u)$ is a consistent estimator of the selector. Note
that we do not need to recompute $\widehat \gamma(u)$ in the weighted
samples, which is convenient for computation.

We make the following assumptions about the estimator of the control
variable in the weighted sample.

\begin{condition}[Weighted estimator of control variable]\label{condition:controlb}
Let $(e_1, \ldots, e_n)$ be a sequence of weights that satisfies
Assumption \ref{condition:weights}. We have an estimator of the control variable of the
form $\widehat{V}^e=\widehat{\vartheta}^e(D,W, Z)$ such that uniformly over
$\overline{\D\W\Z}$,
\begin{equation*}
\sqrt{n}(\widehat{\vartheta}^e(d,w,z)-\vartheta_0(d,w,z))= \frac{1}{\sqrt{n}}%
\sum_{i=1}^{n} e_i \ell(A_i, d,w,z) + o_{\Pr}(1), \ \ \Ep_P[\ell(A,
d,w,z)]=0,
\end{equation*}%
where $\Ep_P[T \ell(A, D,W,Z)^2] < \infty$ and $\| \frac{1}{\sqrt{n}}%
\sum_{i=1}^{n} e_i \ell(A_i, \cdot)\|_{T,\infty} = O_{\Pr}(1)$, and
$$
\| \widehat \vartheta^e -  \widetilde \vartheta^e\|_{T,\infty} =
o_{\Pr}(1/\sqrt{n}), \ \ \text{ for } \ \  \widetilde \vartheta^e
\in \Upsilon
$$
with probability approaching one, where the covering entropy of the function class $\Upsilon$ is not too high, namely
$$
\log N (\epsilon, \Upsilon, \|\cdot\|_{T,\infty}) \lesssim
1/(\epsilon \log^4 (1/ \epsilon)), \text{ for  all } 0 < \epsilon <
1.
$$
\end{condition}

Basically this is the same condition as Assumption \ref{condition:control} in the
unweighted sample, and therefore both can be verified using
analogous arguments. Note also that the condition is stated under
the probability measure $\Pr$, i.e. unconditionally on the data,
which actually simplifies verification. We give primitive conditions
that verify this assumption for quantile and distribution regression
estimation of the control variable below.

 The following result shows the consistency of
weighted bootstrap to approximate the asymptotic distribution of the
CQIV estimator.
\begin{theorem}[Weighted-bootstrap validity for CQIV]
Under the stated assumptions, conditionally on the data
\begin{equation*}
\sqrt{n}(\widehat\beta^{e} (u) - \widehat \beta(u))  \to_d N(0,
J^{-1}(u) \Lambda(u) J^{-1}(u)),
\end{equation*}
in probability under $\Pr$.
\end{theorem}

Note that the statement above formally means that the distance
between the law of $\sqrt{n}(\widehat\beta^{e} (u) - \widehat
\beta(u))$ conditional on the data and the law of the normal vector
$N(0, J^{-1}(u) \Lambda(u) J^{-1}(u))$, as measured by any metric
that metrizes weak convergence, conveges in probability to zero.
More specifically,
$$
d_{BL}\{ \mathcal{L}[\sqrt{n}(\widehat\beta^{e} (u) - \widehat
\beta(u))| \text{data}],  \mathcal{L}[N(0, J^{-1}(u) \Lambda(u)
J^{-1}(u))] \} \to_{\Pr} 0,
$$
where $d_{BL}$ denotes the bounded Lipshitz metric.

In practice, we approximate
numerically the distribution of $ \sqrt{n}(\widehat\beta^{e} (u) -
\widehat \beta(u))$ conditional on the data by simulation. For $b =
1,\ldots,B,$ we compute $\widehat\beta^{e}_{b} (u)$ solving the
problem (\ref{eq: weighted cqiv}) with the data fixed and a set of
weights $(e_{1b}, ..., e_{nb})$ randomly drawn from a distribution
that satisfies Assumption \ref{condition:weights}. By Theorem 2, we can use the empirical
distribution of $\sqrt{n}(\widehat\beta^{e}_b (u) - \widehat
\beta(u))$  to make asymptotically valid inference on $\beta_0(u)$
for large $B$.

\subsection{Quantile and distribution regression estimation of the control variable}

One of the main contributions of this paper is to allow for quantile
and distribution regression estimation of the control variable. The
difficulties here are multifold, since the control variable  depends
on the infinite dimensional function $\pi_0(\cdot)$, and more
importantly the estimated version of this function, $\widehat
\pi(\cdot)$, does not seem to lie in any class with good entropic
properties. We overcome these difficulties by demonstrating that the
estimated function can be approximated with sufficient degree of
accuracy by a random function that lies in a class with good
entropic properties. To carry out this approximation, we smooth the
empirical quantile regression and distribution regression processes
by third order kernels, after suitably extending the processes to
deal with boundary issues. Such kernels  can be obtained by
reproducing kernel Hilbert space methods or via twicing kernel
methods (Berlinet, 1993, and Newey, Hsieh, and Robins, 2004). In the
case of quantile regression, we also use results of the asymptotic
theory for rearrangement-related operators developed by
Chernozhukov, Fern\'andez-Val and Galichon (2010). Moreover, all the
previous arguments carry over weighted samples, which is relevant
for the bootstrap.

\subsubsection{Quantile regression} We impose the following
condition:

\begin{condition}[QR control variable]\label{Ass: QR} (a) The conditional
quantile function of $D$ given $(W,Z)$ follows the quantile
regression model:$$Q_D(v \mid W,Z) = Q_D(v \mid R) =
R'\pi_0(v),  \ \ R = r(W,Z),$$ for all $v \in  \mathcal{T}= [\tau, 1- \tau]$, for some $\tau>0$, where $ \{ F_D(d \mid w,z) : (d,w,z) \in \overline{\D\W\Z}\} \subseteq \mathcal{T}$, and the coefficients $v \mapsto
\pi_0(v)$ are three times continuously differentiable with uniformly
bounded derivatives on $v \in  \mathcal{T}$; (b) $\overline{\R}$ is compact; (c) the
conditional density $f_{D}(d \mid r)$ exists, is uniformly continuous in $(d,r)$ over $\overline{\D\R},$  and is uniformly bounded; and (d) the minimal eigenvalue of $\Ep_P \left[
f_{D}(R'\pi_0(v) \mid R) R R'\right]$ is bounded away from zero uniformly over $v \in \mathcal{T}.$
\end{condition}

We impose that $ \{ F_D(d \mid w,z) : (d,w,z) \in \overline{\D\W\Z}\} \subseteq \mathcal{T}$ to ensure that the untrimmed observations are not at the tails of the conditional distribution, restricting the support of the control variable for these observations, i.e., $\overline{\V} \subseteq \mathcal{T}$. The differentiability  of $v \mapsto
\pi_0(v)$  is used in the proofs to construct a smooth approximation to the quantile regression process.  The rest of the conditions are standard in quantile regression models (see, e.g., Koenker, 2005).

 For $\rho_v(z):= (v - 1(z<0))z$ and $v \in \mathcal{T}$, let
$$
\widehat \pi^e(v) \in \arg \min_{\pi \in \mathbb{R}^{\dim(R)}}
\frac{1}{n} \sum_{i=1}^n e_i \rho_v(D_i - R_i'\pi),
$$
where either  $e_i=1$ for the unweighted sample, to obtain the
estimates; or $e_i$ is drawn from a positive random variable with
unit mean and variance for the weighted sample, to obtain bootstrap
estimates.  Then set
 $$
 \vv_0(d,r) =   \tau + \int_{\T} 1\{r'\pi_0(v) \leq d\} d v ;  \ \widehat \vv^e (d,r) = \tau + \int_{\T} 1\{r'\widehat \pi^e(v) \leq d\} d v,
 $$
if $(d,r) \in \overline{\D\R}$ and $\vv_0(d,r) =\tau$ otherwise.

The following result verifies that our main high-level conditions
for the control variable estimator in Assumptions  \ref{condition:control} and \ref{condition:controlb} hold under
Assumption \ref{Ass: QR}. The verification is done simultaneously for weighted
and unweighted samples by including weights that can be equal to the
trivial unit weights, as mentioned above.

 \begin{theorem}[Validity of Assumptions \ref{condition:control} \& \ref{condition:controlb} for QR]   Suppose that Assumptions \ref{condition: trimming} and \ref{Ass: QR} hold. Then,
 (1)
 \begin{eqnarray*}
\sqrt{n}(\widehat \vv^e(d,r) - \vv_0(d,r)) &=&   \frac{1}{\sqrt{n}} \sum_{i=1}^n e_i \ell(A_i, d,r) + o_\Pr(1) \rightsquigarrow \Delta^e(d,r) \text{ in } \ell^{\infty}(\overline{\mathcal{DR}}), \\
 \ell(A, d,r) &:=&   f_{D}(d \mid r)  r' \Ep_P \left[
f_{D}(R'\pi_0(\vv_0(d,r)) \mid R) R R'\right]^{-1} \times\\
&& \times  [ 1\{D \leq R'\pi_0(\vv_0(d,r))\}- \vv_0(d,r)] R,   \\
 \Ep_P [\ell(A, d,r)] &=& 0, \ \Ep_P [T \ell(A, D,R)^2] < \infty,
 \end{eqnarray*}
where  $\Delta^e(d,r)$ is a Gaussian process with continuous paths
and covariance function given by $\Ep_P[\ell(A, d,r)\ell(A,
\tilde{d},\tilde{r})']$. (2) Moreover, there exists $\widetilde
\vv^e: \overline{\mathcal{DR}} \mapsto [0,1]$  that obeys the same first order
representation uniformly over $\overline{\mathcal{DR}}$, is close to $\widehat \vv^e$ in the sense that
$\| \widetilde \vv^e - \widehat \vv^e\|_{T,\infty}=o_\Pr(1/\sqrt{n})$,
and, with probability approaching one, belongs to a bounded function
class $\Upsilon$  such that
$$
\log N (\epsilon, \Upsilon, \|\cdot\|_{T,\infty}) \lesssim \epsilon^{-1/2},  \ \  0 < \epsilon <1.
$$
Thus, Assumption \ref{condition:control} holds  for the case $e_i=1$, and Assumption \ref{condition:controlb}
holds for the case of $e_i$  being drawn from a positive random
variable with unit mean and variance as in Assumption \ref{condition:weights}. Thus, the
 results of Theorem 1 and 2 apply for the QR estimator of the control variable.
\end{theorem}

\subsubsection{Distribution regression}  We impose the following
condition:

\begin{condition}[DR control variable]\label{condition:DR}
(a) The conditional distribution function of $D$ given $(W,Z)$
follows the distribution regression model, i.e., $$F_D(d \mid
W,Z) = F_D(d \mid R) = \Lambda(R'\pi_0(d)),   \ \ R =
r(W,Z),$$ for all $d \in \overline{\D}$, where $\Lambda$ is either the probit or logit link
function, and the coefficients $d \mapsto \pi_0(d)$ are three times
continuously differentiable with uniformly bounded derivatives; (b)
$\overline{\R}$ is compact; (c) the minimum eigenvalue of
$$
\Ep_P \left[
\frac{\partial \Lambda(R'\pi_0(d))^2}{\Lambda(R'\pi_0(d))[1- \Lambda(R'\pi_0(d))]} RR' \right]
$$
is bounded away from zero uniformly over $d \in \overline{\D},$ where $\partial \Lambda$ is the derivative of $\Lambda$.
\end{condition}
The differentiability  of $d \mapsto
\pi_0(d)$  is used in the proofs to construct a smooth approximation to the distribution regression process.  The rest of the conditions are standard in distribution regression models (see, e.g., Chernozhukov, Fernandez-Val, and Melly, 2013).

For  $d \in \overline{\D}$, let
$$
\widehat \pi^e(d) \in \arg \min_{\pi \in \mathbb{R}^{\dim(R)}}
\frac{1}{n} \sum_{i=1}^n e_i \{ 1(D_i \leq d ) \log \Lambda( R_i'\pi
) + 1(D_i>d) \log [1- \Lambda( R_i'\pi )]\},
$$
where either  $e_i=1$ for the unweighted sample, to obtain the
estimates; or $e_i$ is drawn from a positive random variable with
unit mean and variance for the weighted sample, to obtain bootstrap
estimates. Then set
 $$
 \vv_0(d,r) =  \Lambda(r'\pi_0(d));   \ \widehat \vv^e (d,r) =  \Lambda(r'\widehat \pi^e(d)),
 $$
 if $(d,r) \in \overline{\D\R},$ and $\vv_0(d,r) = \widehat \vv^e (d,r) = 0$ otherwise.

The following result verifies that our main high-level conditions
for the control variable estimator in Assumptions \ref{condition:control} and \ref{condition:controlb} hold under
Assumption \ref{condition:DR}. The verification is done simultaneously for weighted
and unweighted samples by including weights that can be equal to the
trivial unit weights.

 \begin{theorem}[Validity of Assumptions \ref{condition:control} \& \ref{condition:controlb} for DR]   Suppose that Assumptions \ref{condition: trimming} and  \ref{condition:DR} hold.  Then, (1)
 \begin{eqnarray*}
\sqrt{n}(\widehat \vv^e(d,r) - \vv_0(d,r)) &=&   \frac{1}{\sqrt{n}} \sum_{i=1}^n e_i \ell(A_i, d,r) + o_\Pr(1) \rightsquigarrow \Delta^e(d,r) \text{ in } \ell^{\infty}(\overline{\mathcal{DR}}), \\
\ell(A,d,r) &:=&   \partial \Lambda(r'\pi_0(d))  r' \Ep_P \left[
\frac{\partial \Lambda(R'\pi_0(d))^2}{\Lambda(R'\pi_0(d))[1
- \Lambda(R'\pi_0(d))]} RR' \right]^{-1} \times \\
&& \times \frac{1 \{D \leq d\} - \Lambda(R'\pi_0(d)) }{
\Lambda(R'\pi_0(d))[1- \Lambda(R'\pi_0(d))]} \partial
\Lambda(R'\pi_0(d))R, \\  \Ep_P [\ell(A,d,r)] &=& 0, \Ep_P
[T \ell(A,D,R)^2] < \infty,
 \end{eqnarray*}
where  $\Delta^e(d,r)$ is a Gaussian process with continuous paths
and covariance function given by $\Ep_P[\ell(A, d,r)\ell(A,
\tilde{d},\tilde{r})']$.
(2) Moreover, there exists $\widetilde \vv^e:
\overline{\mathcal{DR}} \mapsto [0,1]$  that obeys the same first order
representation uniformly over $\overline{\mathcal{DR}}$, is close to $\widehat \vv^e$ in the sense that
$\|\widetilde \vv^e - \widehat \vv^e\|_{T,\infty}=o_\Pr(1/\sqrt{n})$
and, with probability approaching one, belongs to a bounded function
class $\Upsilon$ such that
$$
\log N (\epsilon, \Upsilon, \|\cdot\|_{T,\infty}) \lesssim
\epsilon^{-1/2}, \ \  0 < \epsilon < 1.
$$
Thus, Assumption \ref{condition:control} holds  for the case $e_i=1$, and Assumption \ref{condition:controlb}
holds  for the case of $e_i$  being drawn from a positive random
variable with unit  mean  and variance as in Assumption \ref{condition:weights}. Thus, the
 results of Theorem 1 and 2 apply for the DR estimator of the control variable.
\end{theorem}

\section{Computation} \label{computation} This section describes the
numerical algorithms to compute the CQIV estimator and weighted
bootstrap confidence intervals.

\subsection{CQIV Algorithm}

The algorithm to obtain CQIV estimates is similar to Chernozhukov
and Hong (2002). We add an initial step to estimate the control
variable $V$. We number this step as 0 to facilitate comparison
with the Chernozhukov and Hong (2002) 3-Step CQR algorithm. \

\medskip

\begin{algorithm}[CQIV]\label{alg: cqiv}
For each desired quantile $u$, perform the following steps: 0) Obtain $\widehat{V}_i = \widehat{\vartheta}(D_i,W_i,Z_i)$
from (\ref{equation: fe_qr_estimator}) or (\ref{equation: fe_dr_estimator}), and construct $\widehat{X}_i=x(D_i,W_i, \widehat{V}_i)$. 1) Select a set of $u$-quantile uncensored observations $J_0(u)=\{i:\Lambda(\widehat{S}_{i}^{\prime }\widehat{\delta
})\geq1-u+k_0(u)\}$,
where  $\Lambda$ is a known link function,  $\widehat S_i = s(\widehat X_i, C_i)$, $s$ is a vector of transformations, $k_0(u)$ is a cut-off such that $0 < k_0(u) < u,$ and $\widehat{\delta
} = \arg \max_{\delta \in \mathbb{R}^{\dim(S)}} \sum_{i=1}^n T_i \{1(Y_i > C_i) \log \Lambda(\widehat{S}_{i}^{\prime } \delta) + 1(Y_i = C_i) \log [1 - \Lambda(\widehat{S}_{i}^{\prime } \delta)] \}, $ where  $T_i = 1(D_i \in \overline{\D})$. 2) Obtain the 2-step CQIV coefficient estimates:  $
\widehat{\beta }^0(u) = \arg \min_{\beta \in \mathbb{R}^{\dim(X)}} \sum_{i \in J_0(u)} T_i \rho _{u}(Y_{i}-%
\widehat{X}_{i}^{\prime }\beta),$ and update the set of $u$-quantile uncensored observations, $
J_1(u)=\{i:\widehat{X}_{i}^{\prime }\widehat{\beta }^0(u)\geq C_{i} +
\varsigma_{1}(u)\}.$ 3) Obtain the 3-step CQIV coefficient estimates $%
\widehat{\beta }^1(u)$, solving the same minimization program as in step 2 with $J_0(u)$ replaced by $J_1(u)$. 4. (Optional)
Update the set of $u$-quantile uncensored observations $J_{2}$ replacing $\widehat{\beta }^0(u)$ by
$\widehat{\beta }^1(u)$ in the expression for $J_1(u)$ in step 2, and iterate this and
the previous step a bounded number of times.

\end{algorithm}


%


\noindent \textbf{Remark 1} (Step 1). We  can obtain $J_0(u)$ with
 a probit, logit, or any other model for the conditional probability of censoring capable of discriminating a subset of $u$-quantile uncensored  observations. For example, 
we can use a logit model with $s(\widehat X_i,C_i)$ including powers or b-splines of the components of $(\widehat X_i,C_i)$ and interaction terms. Given the slackness provided by the
cut-off $k_0(u)$, the model does not need
to be correctly specified. It suffices to select a nontrivial
subset of observations with $X_i'\beta_0(u)>C_i$. To choose the
value of $k_0(u)$, it is advisable that a constant fraction of
observations satisfying $\Lambda(\widehat{S}_{i}^{\prime
}\widehat{\delta })>1-u$ are excluded from $J_0(u)$ for each
quantile. To do so, set $k_0(u)$ as the
$q_{0}$th quantile of $\Lambda(\widehat{S}_{i}^{\prime }\widehat{%
\delta })$ conditional on $\Lambda(\widehat{S}_{i}^{\prime
}\widehat{\delta})>1-u$, where $q_{0}$ is a percentage (10\% worked
well in our simulation with little sensitivity to values between 5 and 15\%).

\medskip

\noindent \textbf{Remark 2} (Step 2). To choose the cut-off
$\varsigma_{1}(u)$, it is advisable that a constant
fraction of observations satisfying $\widehat{X}_{i}^{\prime }\widehat{%
\beta }^0(u)>C_{i}$ are excluded from $J_1(u)$ for each quantile. \
To do so,
set $\varsigma_{1}(u)$ to be the $q_{1}$th quantile of $\widehat{X}%
_{i}^{\prime }\widehat{\beta }^0(u) - C_i$ conditional on $\widehat{X}_{i}^{\prime }%
\widehat{\beta }^0(u)>C_{i}$, where $q_{1}$ is a percentage less
than $q_{0}$
(3\% worked well in our simulation with little sensitivity to values between 1 and 5\%). In practice, it is desirable that $%
J_0(u)$ $\subset $ $J_1(u)$. If this is not the case, we recommend
altering $q_{0}$, $q_{1}$, or the specification of the regression
models. At each quantile,
the percentage of observations from the full sample retained in $J_0(u)$,
 the percentage of observations from the full sample retained in $J_1(u)$,
 and the percentage of observations from $J_0(u)$ not retained in $J_1(u)$
can be computed as simple robustness diagnostic tests. The estimator $%
\widehat{\beta }^0(u)$ is consistent but will be less inefficient than the estimator obtained in the subsequent step because it uses a
smaller conservative subset of the $u$-quantile uncensored observations if $q_1 < q_0$.


\medskip

\noindent \textbf{Remark 3} (Steps 1 and 2).  In the notation of
Assumption \ref{condition: selector}, the selector of Step 1 can be
expressed as $1(\widehat{S}_i'\widehat{\gamma}(u) \geq \varsigma_0(u))$,
where $\widehat{S}_i'\widehat{\gamma}(u) =
\widehat{S}_i'\widehat{\delta} - \Lambda^{-1}(1-u)$ and $\varsigma_0(u)
= \Lambda^{-1}(1-u+k_0(u)) - \Lambda^{-1}(1-u)$. The selector of Step 2
can  be expressed as $1(\widehat{S}_i'\widehat{\gamma}(u) \geq
\varsigma_1(u)),$ where $\widehat{S}_i = (\widehat{X}_i', C_i)'$ and
$\widehat{\gamma}(u) = (\widehat{\beta}^{0}(u)',-1)'$.

\medskip

\noindent \textbf{Remark 4} (Steps 1 and  2). The trimming rule is a useful theoretical device that is generally considered to have minor practical importance. In our numerical and empirical examples,  the choice of $\overline{\D}$ as the observed support of $D,$ i.e. no trimming, works well. We also found that the performance of the estimator is not sensitive to the use of other trimming rules such as $\overline{\D} = [\widehat{D}_{\tau},\widehat{D}_{1-\tau}]$ where $\widehat{D}_{\tau}$ is the empirical $\tau$-quantile of $D$ for some small $\tau$ (e.g, $\tau = .01$).

\medskip



\noindent \textbf{Remark 5} (Steps 2, 3 and 4). The CQIV  algorithm
provides a computationally convenient approximation to Powell's censored quantile regression estimator.  As a
simple robustness diagnostic test, we recommend computing  the
Powell objective function using the full sample and the estimated
coefficients after each iteration, starting with Step 2. This
diagnostic test is computationally straightforward because computing
the objective function for a given set of values is much simpler
than maximizing it. In practice, this test can be used to determine
when to stop the CQIV algorithm for each quantile. If the Powell
objective function increases from Step $s$ to Step $s+1$ for $s\geq
2$, estimates from Step $s$ can be retained as the coefficient
estimates.

\medskip

\noindent \textbf{Remark 6} (Step 4).  Iterating over the 3-step CQIV estimator with fixed cutoff at $\varsigma_1(u)$ does not affect asymptotic efficiency, but it might improve finite-sample properties.  In our numerical experiments, however, we find very little or no gain of iterating beyond Step 3 in terms of bias, root mean square error, and value of Powell objective function.

\subsection{Weighted Bootstrap Algorithm}

We recommend obtaining confidence intervals through a weighted
bootstrap procedure, though analytical formulas can also be used. If
the estimation runs quickly on the desired sample, it is
straightforward to rerun the entire CQIV algorithm $B$ times
weighting all the steps by the bootstrap weights. To speed up the
computation, we propose a procedure that uses a one-step CQIV
estimator in each bootstrap repetition.

\begin{algorithm}[Weighted bootstrap CQIV] \label{alg: bootstrap} For $b = 1, \ldots, B$, repeat the following steps:
1) Draw a set of weights $(e_{1b}, \ldots, e_{nb})$
i.i.d from the standard exponential distribution or another distribution that satisfies Assumption \ref{condition:weights}. 2) Reestimate the control variable in the weighted sample,  $\widehat V_{ib}^{e} = \widehat \vartheta_{b}^e(D_i, W_i,
Z_i)$, and construct $\widehat X_{ib}^{e} = x(D_i, W_i, \widehat
V_{ib}^{e})$. 3) Estimate the weighted quantile regression:
$
\widehat{\beta }_{b}^{e}(u) = \arg \min_{\beta \in \mathbb{R}^{\dim(X)}} \sum_{i \in J_{1b}} e_{ib} T_i \rho _{u}(Y_{i}-%
\beta^{\prime}\widehat{X}_{ib}^{e}),  \label{QR}
$
where $J_{1b} = \{i:\widehat{\beta }(u)^{\prime}\widehat{X}_{ib}^{e}
\geq C_{i} + \varsigma_{1}(u) \},$ and $\widehat{\beta }(u)$ is a consistent estimator
of $\beta_0(u)$, e.g., the 3-stage CQIV estimator $\widehat
\beta^1(u)$.

\end{algorithm}

\medskip

%

\noindent \textbf{Remark 7} (Step 3). A computationally less
expensive alternative is to set $J_{1b} = J_{1}$ in all the
repetitions, where $J_1(u)$ is the subset of selected observations in
Step 2 of the CQIV algorithm.

\medskip

We can construct an asymptotic $(1-\alpha)$-confidence interval for
a scalar function of the parameter vector $g(\beta_0(u))$ using the percentile method, i.e., $CI_{1-\alpha}[g(\beta_0(u))] = [\widehat
g_{\alpha/2}, \widehat g_{1-\alpha/2}]$, where $\widehat g_{\alpha}$
is the sample $\alpha$-quantile of $[g(\widehat \beta^{e}_{1}(u)),
\ldots, g(\widehat \beta^{e}_{B}(u))]$. For example, let $\widehat{\beta }_{b,k}^{e}(u)$ and $\beta_{0,k}(u)$
denote the $k$th components of $\widehat{\beta }_{b}^{e}(u)$ and $\beta_{0}(u)$. Then, the 0.025 and
0.975 quantiles of $(\widehat{\beta }_{1, k}^{e}(u), \ldots,
\widehat{\beta }_{B,k}^{e}(u))$ form a 95\% asymptotic confidence
interval for  $\beta_{0,k}(u)$.


\section{Monte-Carlo illustration}

\label{montecarlo}

In this section, we develop a Monte-Carlo numerical example aimed at
analyzing the performance of CQIV in finite samples.  We first generate
data according to two different designs.  Next,
we compare the performance of CQIV and tobit estimators in terms of bias and root mean squared error.
Finally, we discuss the results of sensitivity and diagnostic tests calculated within the
simulated data.

\subsection{Monte-Carlo Designs}

We generate data according to a design that satisfies the
tobit parametric assumptions and a design with heteroskedasticity in
the first stage equation for the endogenous regressor $D$ that does
not satisfy one of  the tobit parametric assumptions.\footnote{The tobit parametric assumptions are classical location models for  the first stage and second stage  equations and jointly normal unobservables.} To facilitate the
comparison, in both designs we consider a location model for the
latent variable $Y^{\ast}$, where the coefficients of the
conditional expectation function and the conditional quantile
function are equal (other than the intercept), so that tobit and
CQIV estimate the same parameters.
A comparison of the dispersion of the tobit estimates to the
dispersion of the CQIV estimates at each quantile in the first
design serves to quantify the relative efficiency of CQIV in a case
where tobit can be expected to perform as well as possible.

For the tobit design, we use the following simplified
version of the  system of equations (\ref{eq: normal1})--(\ref{eq:
normal2}) to generate the observations:
 \begin{eqnarray}
D &=& \pi_{00}+ \pi_{01}Z+ \pi_{02}W+\Phi ^{-1}(V),\text{ \ \ }%
V\backsim U(0,1),\text{ \ \ }  \label{sim first stage} \\
Y^{\ast } &=& \beta_{00}+ \beta_{01}D+\beta_{02}W+\Phi
^{-1}(\epsilon),\text{ \ \ }\epsilon \backsim U(0,1), \label{sim
structural} \end{eqnarray} where $\Phi^{-1}$ denotes the quantile
function of the standard normal distribution, and $(\Phi ^{-1}(V)$,
$\Phi ^{-1}(\epsilon))$ is jointly normal with correlation $\rho_0$.
\
Though we can observe $Y^{\ast }$ in the simulated data, we
artificially
censor the data to
\begin{equation} Y = \max (Y^{\ast },C) = \max (\beta_{00}+
\beta_{01}D+\beta_{02}W+\Phi ^{-1}(\epsilon),C). \label{sim
structural censored} \end{equation} From the properties of the
multivariate normal distribution, $\Phi ^{-1}(\epsilon)= \rho_0 \Phi
^{-1}(V)+(1 - \rho_0^2)^{1/2} \Phi ^{-1}(U)$, where $U \backsim
U(0,1)$. \ Using this expression, we can combine (\ref{sim structural}) and (\ref{sim structural censored}) for
an alternative formulation of the censored model in which the control term $%
V_i $ is included in the  equation for the observed response:
\begin{equation*} Y = \max (Y^{\ast },C) = \max (\beta_{00}+
\beta_{01}D+\beta_{02}W+\rho_0 \Phi ^{-1}(V)+ (1 - \rho_0^2)^{1/2}
\Phi ^{-1}(U),C). \end{equation*} This formulation is useful because
it indicates that when we include the control variable in the quantile
function, its true coefficient is $\rho_0 $.

In our simulated data, we create extreme endogeneity by setting
$\rho_0 =.9$. We set $\pi _{00}= \beta _{00}=0$, and $\pi_{01} =
\pi_{02} = \beta _{01}= \beta_{02}=1$. We draw the disturbances
$[\Phi ^{-1}(V),\Phi ^{-1}(\epsilon)]$ from a
bivariate normal distribution with zero means, unit variances and correlation $\rho_0$. We draw $%
Z $ from a standard normal distribution, and we generate $W$ to be a
log-normal random variable that is censored from the right at its
95th percentile. Formally, we set\ $W=\exp[\min
(W^*,q_{W^*})],$ where $W^*$ is drawn from a
standard normal distribution and $q_{W^*}$ is the 95th sample percentile of $W^*$,
which differs across replication samples. For comparative purposes, we set the
amount of censoring in the dependent
variable to be comparable to that in Kowalski (2009). Specifically, we set $C$ to the 38th sample percentile of
$Y^*$ in each replication sample. We report results
from 1,000 simulations with $n=1,000$.

For the design with heteroskedastic first stage, we replace the first stage equation
for $D$ in (\ref{sim first stage}) by the following equation:
\begin{equation} D = \pi_{00}+ \pi_{01}Z+
\pi_{02}W+(\pi_{03}+\pi_{04}W)\Phi ^{-1}(V),\text{ \ \ }V\backsim
U(0,1)\text{ \ \ } \label{het first stage} \end{equation} where we
set $\pi _{03}=\pi _{04}=1$. The corresponding conditional quantile
function is $$ Q_{D}(v \mid W,Z) =  \pi_{00}+ \pi_{01}Z+
\pi_{02}W+(\pi_{03}+\pi_{04}W)\Phi ^{-1}(v), $$ which can be
consistently estimated by quantile regression or other estimator for
location-scale shift models.


\subsection{Comparison of CQIV with Tobit}

We consider  two  tobit estimators for comparison. Tobit-iv is the
full information maximum likelihood estimator
implemented in Stata with the default option of the command
\verb"ivtobit".\footnote{The results reported use the algorithm ``difficult'' because the default algorithm does not converge in several simulations for the heteroskedastic design. The algorithm ``bfgs'' and the Newey's (1987) minimum chi-squared option of the command give similar results to the ones reported.}
 Tobit-cmle is the conditional maximum likelihood
tobit estimator developed by Smith and Blundell (1986), which uses
least squares residuals as a parametric control variable.
For CQIV we consider three different methods to estimate the control variable: cqiv-ols,
which uses least squares to estimate a  parametric control variable; cqiv-qr, which uses quantile regression to estimate a semiparametric control variable;
and cqiv-dr, which uses probit distribution regression to estimate a semiparametric control variable.\footnote{See appendix for technical details on the computation of the first stage estimators of the control variable.} All the CQIV estimators are computed in three stages using Algorithm \ref{alg: cqiv} with $q_0 = 10,$ $q_1 = 3,$ no trimming, and a probit model in step 1.

\begin{figure}[tbp]
\begin{center}
\includegraphics[width=.9\textwidth, height = .45\textheight, angle=0]{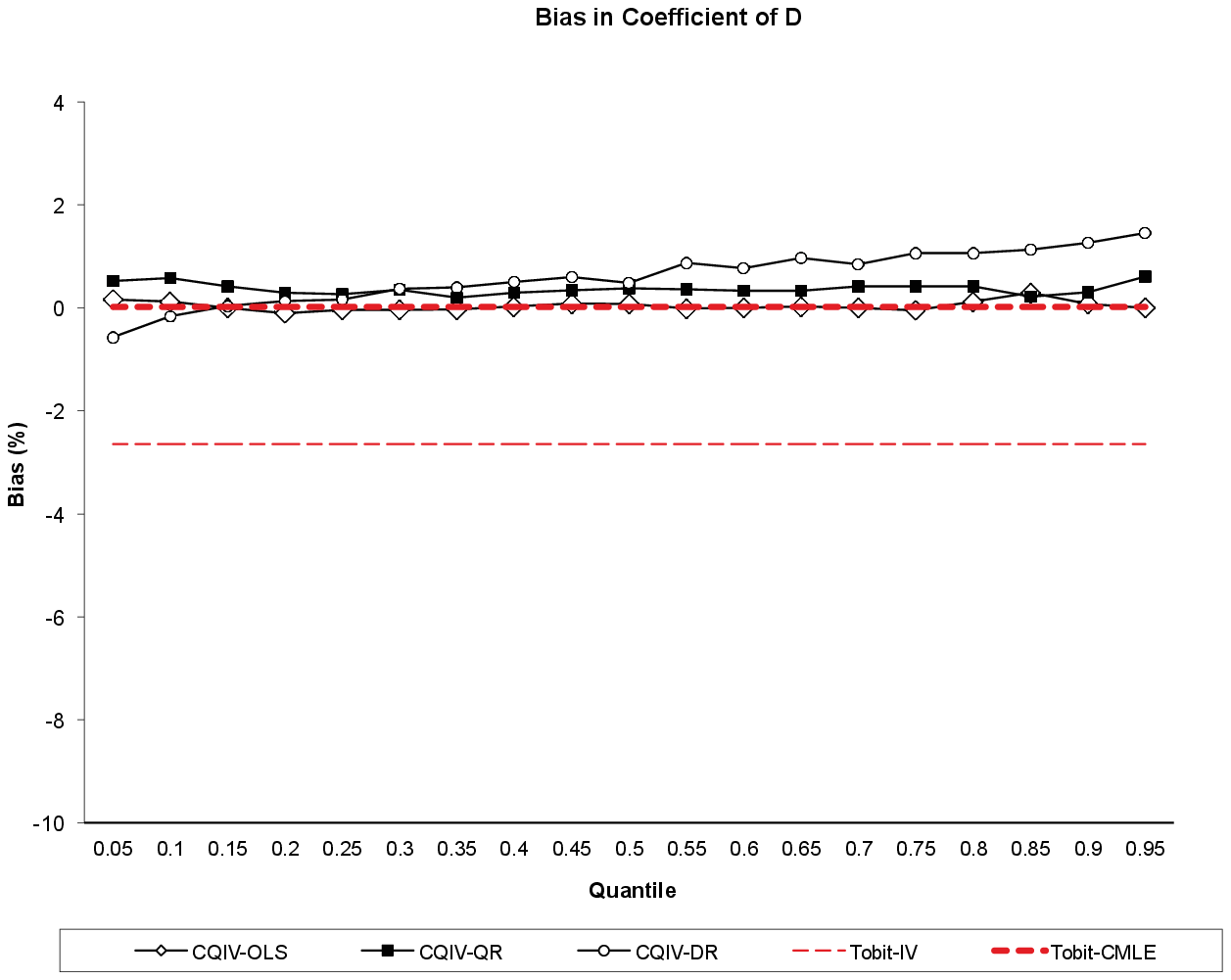}
\includegraphics[width=.9\textwidth, height = .45\textheight, angle=0]{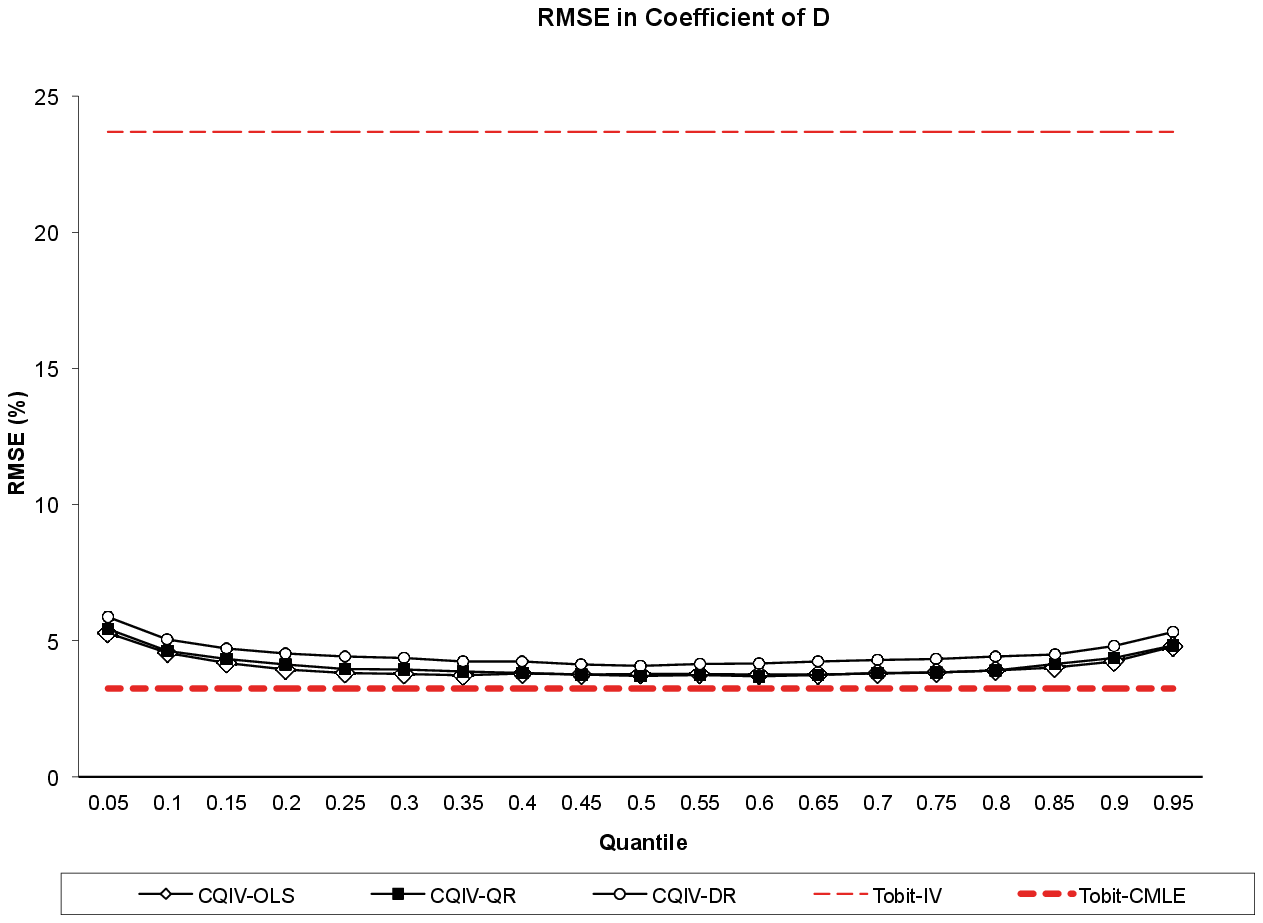}
\end{center}
\caption{Tobit design: Mean bias and rmse of tobit  and cqiv
estimators. Results obtained from 1,000 samples of  size $n =
1,000$.} \label{fig: homos}
\end{figure}

We focus on the coefficient on the endogenous regressor $D$. We
report mean bias and root mean square error (rmse) for all the
estimators at the $\{.05, .10, ...,  .95\}$ quantiles.
 For the tobit design, the bias results are reported in the
upper panel of Figure \ref{fig: homos} and the rmse results are
reported in the lower panel. In this figure, we see that tobit-cmle
represents a substantial improvement over tobit-iv in terms of mean
bias and rmse.
Even though tobit-iv is theoretically asymptotically efficient in this design, the
CQIV estimators out-perform tobit-iv, and compare well to
tobit-cmle.
Cqiv-ols and cqiv-qr display slightly lower rmse than  cqiv-dr in this design. All of our qualitative
findings hold when we consider unreported alternative measures of
bias and dispersion such as median bias, interquartile range, and
standard deviation.

\begin{figure}[tbp]
\begin{center}
\includegraphics[width=.9\textwidth, height = .45\textheight, angle=0]{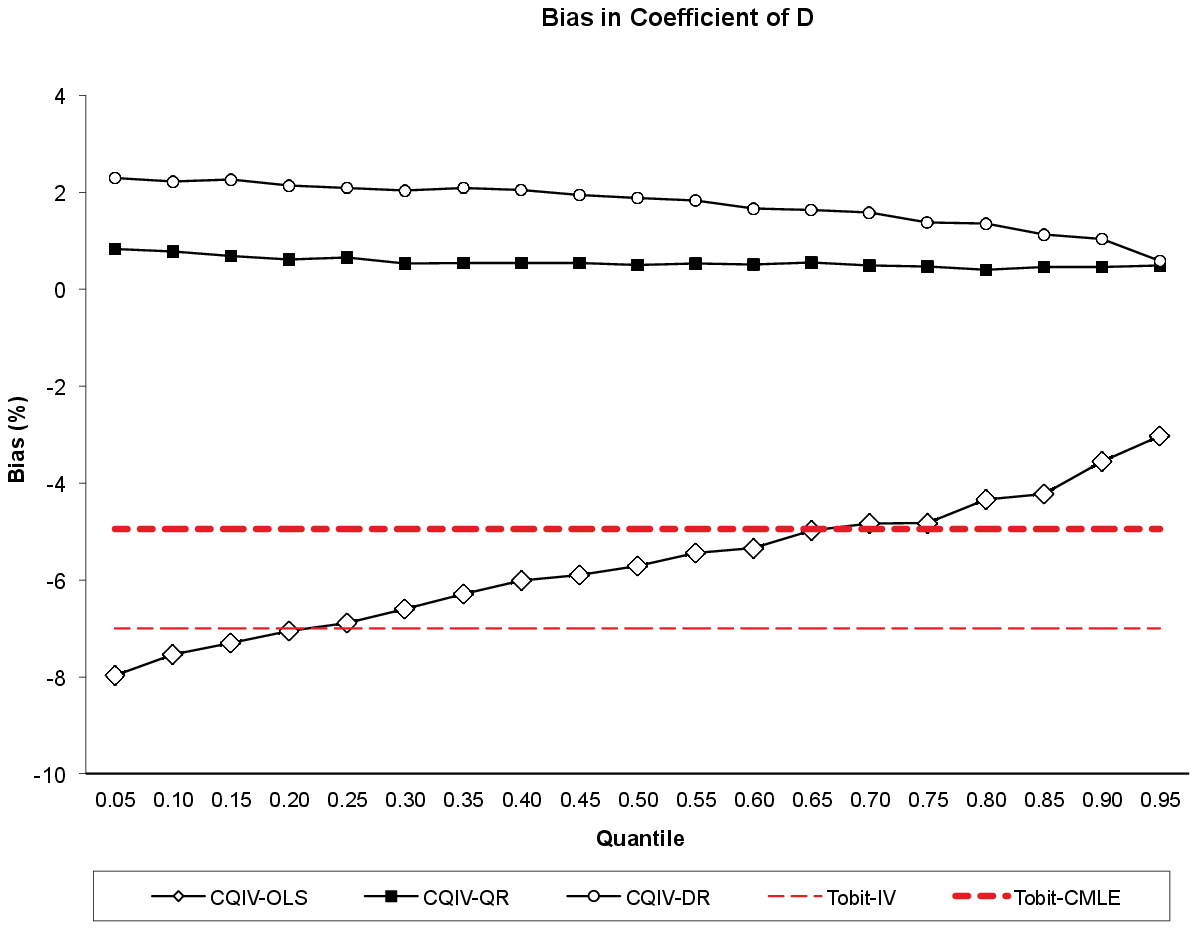}
\includegraphics[width=.9\textwidth, height = .45\textheight, angle=0]{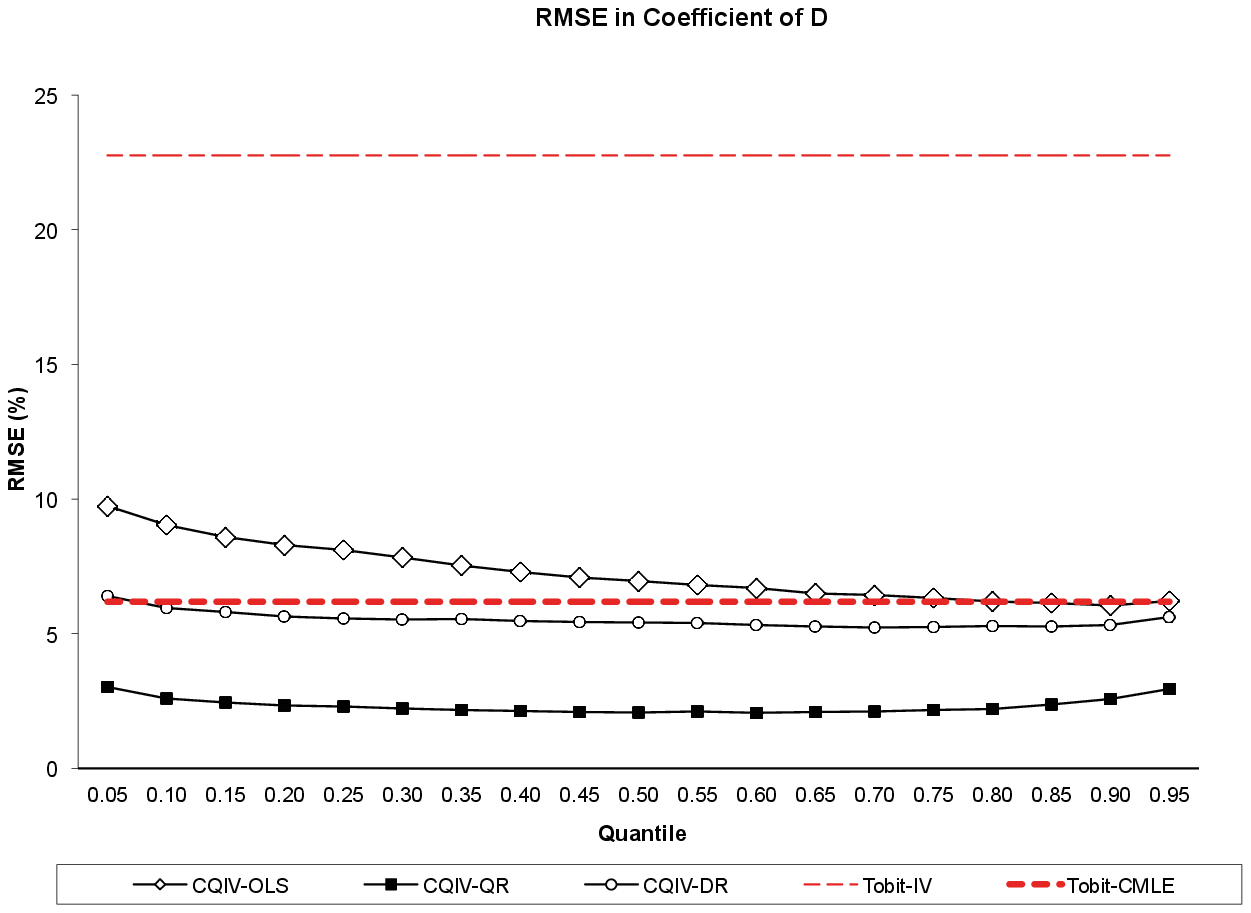}
\end{center}
\caption{Design with heteroskedastic first stage: Mean bias and rmse of tobit  and cqiv
estimators. Results obtained from 1,000 samples of  size $n =
1,000$.} \label{fig: heteros}
\end{figure}

The similar performance of tobit-cmle and cqiv can be explained by
the homoskedasticity in the first stage of the design. Figure
\ref{fig: heteros} reports mean bias and rmse results for the
design with heteroskedastic first stage. Here cqiv-qr outperforms cqiv-ols and
cqiv-dr at every quantile, which is expected because cqiv-ols and
cqiv-dr are both misspecified for the control variable. Cqiv-dr has
lower bias and rmse than cqiv-ols because it uses a more flexible
specification for the control variable.
Moreover, at every quantile, cqiv-qr outperforms both tobit
estimators, which are no longer consistent.

In summary,
CQIV performs well relative to tobit in a model that satisfies the
parametric assumptions required for tobit-iv to be asymptotically efficient, and it
outperforms tobit in a more flexible model that does not satisfy one of the tobit parametric assumptions.

\subsection{Sensitivity and Diagnostic Tests}

In Table 1, we analyze the sensitivity of the CQIV estimator to the
choice of the quantiles $q_0$ and $q_1$ that determine the cut-offs of the selectors. For all the
combination of values of $q_0 \in \{5, 10, 15\}$ and $q_1 \in
\{1,2,5\},$ we report the mean bias and rmse of the 3-step cqiv-qr
estimator in the tobit design  and the design with heteroskedastic first stage. We find
that the performance of the estimator shows very little sensitivity
to the choice of quantiles within the range of values considered. In
results not reported, we also find very little sensitivity to the
choice of quantiles in the value of the Powell objective function
computed from the 3-step estimator.

Table 2 reports feasible and unfeasible diagnostic tests for the
2-step, 3-step, and 4-step cqiv-qr estimators obtained by Algorithm
\ref{alg: cqiv}  with $q_0 = 10,$ $q_1 = 3,$ and $q_2 = 3$ for both
the tobit and nontobit designs.  We recommend that
applied researchers conduct the feasible tests.  The unfeasible
tests are those that involve $J^*(u) = \{ i:  X_i'\beta_0(u) > C_i \},$ the set of $u$-quantile uncensored
observations, that is unobservable in practice. As shown in the
table, the percentage of observations in $J(u)$ increases with the
quantile.  In the table, we compare the composition of  $J(u)$ with the
compositions of $J_0(u)$ and $J_1(u),$ the subsets of observations selected
as $u$-quantile uncensored in the step 1 and step 2 of the algorithm. We
find that $J_0(u)$ and $J_1(u)$ select most of the $u$-quantile uncensored
observations.

The feasible tests are based on calculating the percentage of
observations selected in $J_0(u)$ and $J_1(u)$, comparing the composition
of the subsets $J_0(u)$ and $J_1(u),$ and calculating the value of the
Powell objective function at each step of the algorithm.  We find
that the percentage of observations retained in $J_0(u)$ and $J_1(u)$
increases with the quantile, as it should given the percentage of
observations in $J$. In applied settings, researchers can diagnose a
problem if the number of observations retained in $J_0(u)$ and $J_1(u)$
varies little across quantiles and attempt to address it by making
the specifications of the binary choice model in step 1 or the
quantile regression model in steps 2 and 3 more flexible. We find
that $J_0(u)$ is a strict subset of $J_1(u)$ in the column that reports
the intersection of $J_0(u)$ with the complement of $J_1(u)$ ($J_0(u) \cap
J_1(u)^{c}$).  In applied settings, researchers can diagnose a problem
if many observations from $J_0(u)$ are not included in $J_1(u)$ and
attempt to address it by modifying the values of $q_{0}$ and
$q_{1}$. The value of the Powell objective function decreases
between step 2 and step 3 of the algorithm in about 75\% of the
simulations, whereas it only further decreases with an additional
iteration in about 25\% of the simulations.  In applied settings,
researchers can use the relative values of the Powell objective
function to assess the gains from iteration.

\section{Empirical Application: Engel Curve Estimation}

\label{engel}


In this section, we apply the CQIV estimator to the estimation of
Engel curves. The Engel curve relationship describes how a
household's demand for a commodity changes as the household's
expenditure increases. Lewbel (2006) provides a recent survey of the
extensive literature on Engel curve estimation. For comparability to
the recent studies, we use data from the 1995 U.K. Family
Expenditure Survey (FES) as in Blundell, Chen, and Kristensen (2007)
and Imbens and Newey (2009). Following Blundell, Chen, and
Kristensen (2007), we restrict the sample to 1,655 married or
cohabitating couples with two or fewer children, in which the head
of household is employed and between the ages of 20 and 55. The FES
collects data on household expenditure for different categories of
commodities. We focus on estimation of the Engel curve relationship
for the alcohol category because 16\% of families in the data report
zero expenditure on alcohol. Although zero expenditure on alcohol
arises as a corner solution outcome, and not from bottom coding,
both types of censoring motivate the use of censored estimators such
as CQIV.

Endogeneity in the estimation of Engel curves arises because the
decision to consume a particular category of commodity occurs
simultaneously with the allocation of income between consumption and
savings. Following the literature, we rely on a two-stage budgeting
argument to justify the use of labor income as an instrument for
expenditure (Gorman, 1959). Specifically, we estimate a
quantile regression model in the first stage, where the logarithm of total expenditure, $%
D$, is a function of the logarithm of gross earnings of the head of
the household, $Z$, and demographic household characteristics, $W$.
The control variable, $V$, is obtained using the quantile regression estimator
in (\ref{equation: fe_qr_estimator}), where $\tau = .01$ and the integral is
approximated by a grid of 100 quantiles. For comparison, we also
obtained control variable estimates using least squares and  probit
distribution regression. We do not report these comparison methods
because the correlation between the different control variable
estimates was virtually 1, and all the methods resulted in very
similar estimates in the second stage.

In the second stage we focus on the following quantile specification for
Engel curve estimation:
\begin{equation*}
Y_{i}=\max (X_{i}^{\prime }\beta_0 (U_{i}),0),\
X_{i}=(1,D_{i},D_{i}^{2},W_{i},\Phi^{-1}(V_{i})),\ U_{i}\backsim
U(0,1)\mid X_{i},
\end{equation*}%
where $Y$ is the observed share of total expenditure on alcohol
with a mass point at zero, $W$ is a binary household demographic variable
that indicates whether the family has any children, and $V$ is the
control variable. We define our binary demographic variable
following
Blundell, Chen and Kristensen (2007).\footnote{%
Demographic variables are important shifters of Engel curves. In
recent literature, \textquotedblleft shape invariant" specifications
for demographic variable have become popular. For comparison with
this literature, we also estimate an unrestricted version of shape
invariant specification in which we include a term for the
interaction between the logarithm of expenditure and our demographic
variable. The results from the shape invariant specification are
qualitatively similar but less precise than the ones reported in
this application.}
To choose the specification, we rely on recent studies in Engel
curve estimation. Thus, following Blundell, Browning, and Crawford
(2003) we impose separability between the control variable and other
regressors. Hausman, Newey, and Powell (1995) and Banks, Blundell,
and Lewbel (1997) show that the quadratic specification in
log-expenditure gives a better fit than the linear specification
used in earlier studies. In particular, Blundell, Duncan, and
Pendakur (1998) find that the quadratic specification gives a good
approximation to the shape of the Engel curve for alcohol. To check
the robustness of the specification to the linearity in the control
variable, we also estimate specifications that include nonlinear
terms in the control variable. The results are very similar to the
ones reported.

\begin{figure}[tbp]
\begin{center}
\includegraphics[height= .95\textheight, width = \textwidth, angle=0]{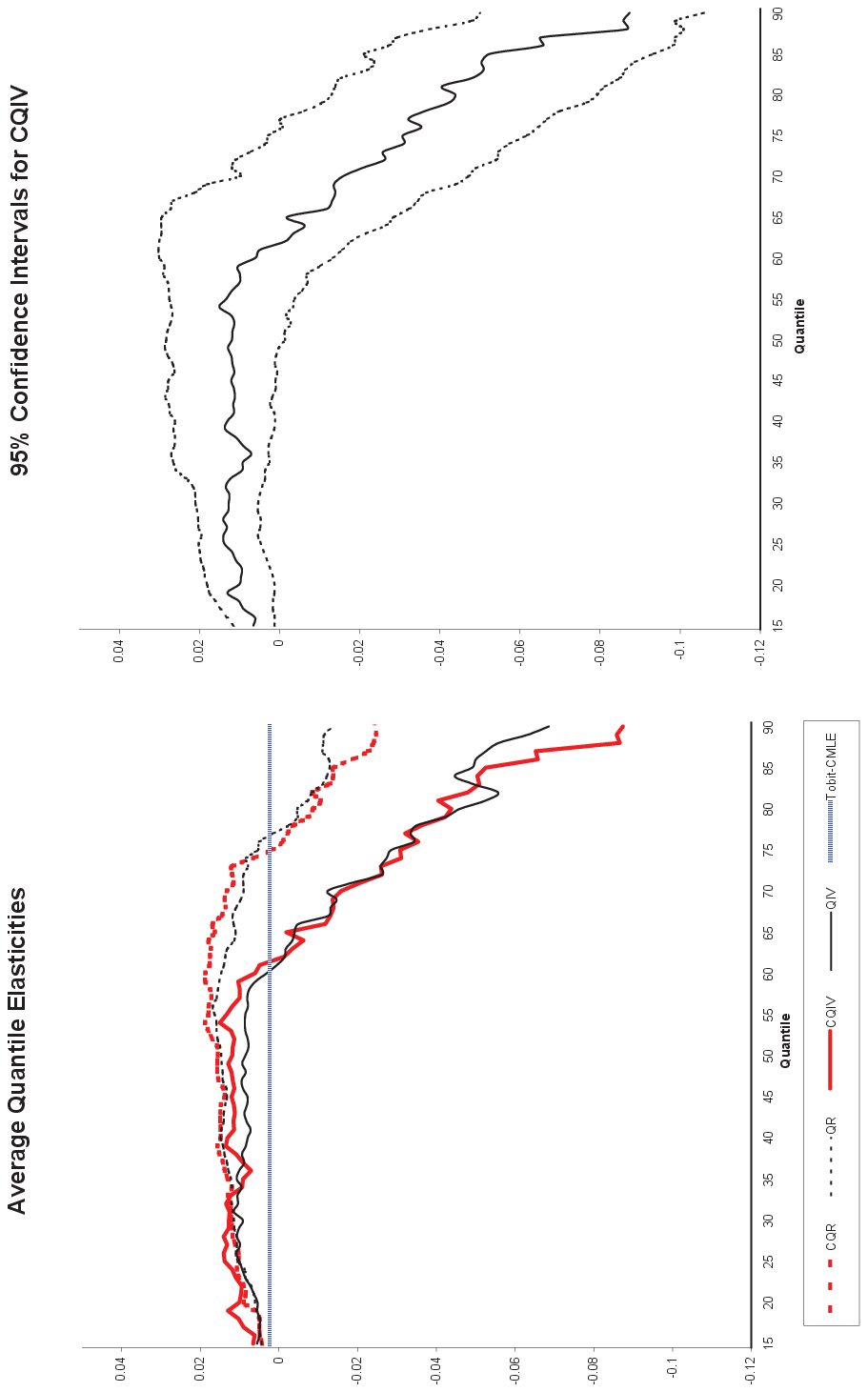}
\end{center}
\caption{Estimates and 95\% pointwise confidence intervals for
average quantile expenditure elasticities. The intervals are
obtained by weighted bootstrap with 200 replications and
exponentially distributed  weights. } \label{elasticities}
\end{figure}

Our quadratic quantile model is flexible in that it permits the
expenditure elasticities to vary across quantiles of the alcohol
share and across the level of total expenditure. These quantile
elasticities are related to the coefficients of the model by
\begin{equation*}
\partial_d Q_Y(u \mid x) =   1\{
x'\beta_0(u) > 0\} \{\beta_{01}(u) + 2 \beta_{02}(u) \ d\},
\label{elasform}
\end{equation*}
where $\beta_{01}(u)$ and $\beta_{02}(u)$ are the coefficients of
$D$ and $D^2$, respectively. Figure \ref{elasticities} reports point
and interval estimates of average quantile elasticities as a
function of the quantile index $u$, i.e., $u \mapsto
\Ep_P[\partial_d Q_Y(u \mid X)]$.   In addition
to CQIV with a quantile estimator of the
control variable (cqiv),   we present results from the censored quantile
regression (cqr) estimator of Chernozhukov and Hong (2002), which
does not address endogeneity; two-stage quantile regression
estimator (qiv)  with quantile regression first
stage, which does not account for censoring; and the quantile
regression (qr) estimator of Koenker and Bassett (1978), which does
not account for endogeneity nor censoring.  We also estimate a model
for the conditional mean with the tobit-cmle of Smith and Blundell
(1986).
Given the level of censoring, we focus
on conditional quantiles above the .15 quantile.

Fig. \ref{elasticities} shows that accounting for
endogeneity and censoring has important consequences for the
elasticities.
The difference between the
estimates is more pronounced along the endogeneity dimension than it
is along the censoring dimension. The right panel plots 95\%
pointwise confidence intervals for the cqiv quantile elasticity
estimates obtained by the weighted bootstrap method described in Algorithm \ref{alg: bootstrap}
 with standard exponential weights and $B=200$ repetitions.
Here we can see that there is significant heterogeneity in the
expenditure elasticity across quantiles. Thus, alcohol passes from
being a normal good for low quantiles to being an inferior good for
high quantiles. This heterogeneity is missed by the tobit
estimates of the elasticity.

\begin{figure}[tbp]
\begin{center}
\includegraphics[width=1\textwidth, height= 1\textheight]{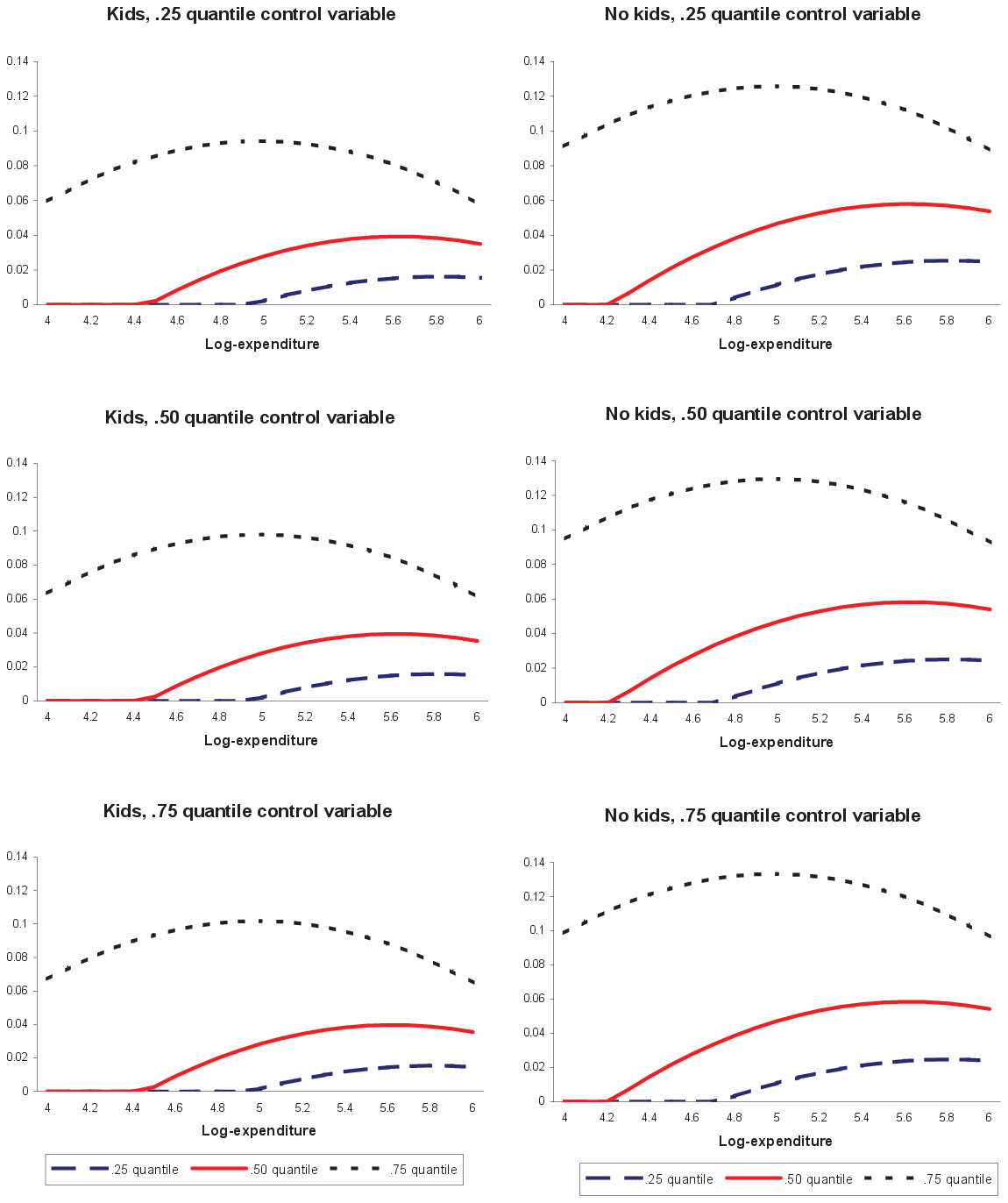}
\end{center}
\caption{Family of Engel curves: each panel plots Engel curves for
the three quantiles of alcohol share.} \label{engelfamily}
\end{figure}

In Figure \ref{engelfamily}\thinspace\ we report families of Engel
curves based on the cqiv coefficient estimates. We predict the value
of the alcohol share, $Y$, for a grid of values of log expenditure
using the cqiv coefficients at each quartile. The subfigures depict
the Engel curves for each quartile of the empirical values of the
control variable, for individuals with and without kids, that is
$$
d \mapsto \max\{(1,d,d^2,w,\Phi^{-1}( v))'\widehat{\beta}(u), 0 \}
$$
for $(w,\Phi^{-1}( v), u)$ evaluated at $w \in \{0,1\}$, the
quartiles of $\widehat V$ for $v$, and $u \in \{0.25, 0.50, 0.75\}$.
Here we can see that controlling for censoring has an important
effect on the shape of the Engel curves even at the median ($u=.5$). The
families of Engel curves are fairly robust to the values of the
control variable, but the effect of children on alcohol shares is
more pronounced. The presence of children in the household produces
a downward shift in the Engel curves at all the levels of
log-expenditure considered.

\bigskip

\section{Conclusion}

\label{conclusion}

In this paper, we develop  new censored and uncensored quantile instrumental
variable estimators that incorporate endogenous regressors using a
control variable approach.  Censoring and endogeneity abound in
empirical work, making the new estimator a valuable addition to the
applied econometrician's toolkit. For example, Kowalski (2009) uses
this estimator to analyze the price elasticity of expenditure on
medical care across the quantiles of the expenditure distribution,
where censoring arises because of the decision to consume zero care
and endogeneity arises because marginal prices explicitly depend on
expenditure. Since the new estimator can be implemented using
standard statistical software, it should prove useful to applied
researchers in many applications.


\appendix

\section{Notation} In what follows $\vartheta$ and $\gamma$ denote
generic values for the control function and the parameter of the
selector $1(S_i'\gamma
\geq \varsigma)$. It is convenient also to introduce some additional notation, which will be
extensively used in the proofs. Let  $V_i(\vv) := \vv(D_i,W_i,Z_i)$, $X_i(\vartheta) := x(D_i, W_i,
V_i(\vartheta))$,  $S_i(\vartheta): =
s(X_i(\vartheta),C_i)$, $\dot{X}_i(\vartheta) :=
\partial_v x(D_i, W_i, v)|_{v = V_i(\vartheta)},$  and $\dot{S}_i(\vartheta) := \partial_v
s(X_i(v),C_i)|_{v = V_i(\vartheta)}$. When the previous
functions are evaluated at the true values we use $V_i = V_i(\vartheta_0),$  $X_i =
X_i(\vartheta_0)$, $S_i = S_i(\vartheta_0)$, $\dot{X}_i =
\dot{X}_i(\vartheta_0)$, and $\dot{S}_i = \dot{S}_i(\vartheta_0)$.
Also, let $\varphi_u(z) := [1(z <0) - u]$. Recall that $ A := (Y,
D,W, Z, C, X, V)$, $T(d) = 1(d \in \overline{\D}),$ and $T = T(D)$.  For a function $f: \mathcal{A} \mapsto \Bbb{R}$,
we use $\| f \|_{T,\infty} = \sup_{a \in \mathcal{A}}|T(d) f(a)|$; for a
$K$-vector of functions $f: \mathcal{A} \mapsto \Bbb{R}^K$, we use
$\|f \|_{T,\infty} = \sup_{a \in \mathcal{A}}\|T(d) f (a) \|_2$.  We make
functions in $\Upsilon$ as well as estimates $\widehat \vartheta$ to
take values in $[0,1]$, the support of the control variable $V$.
This allows us to simplify notation in what follows. We also adopt
the standard notation in the empirical process literature (see,
e.g., van der Vaart, 1998),
$$\En[f] = \En[f(A)] = n^{-1} \sum_{i=1}^n f(A_i),$$ and $$\Gn[f] =
\Gn[f(A)]= n^{-1/2} \sum_{i=1}^n ( f(A_i) - \Ep_P[f(A)] ).$$ When
the function $\widehat f$ is estimated, the notation should
interpreted as:
$$
\Gn[\widehat f \ ] = \Gn[f]\mid_{f = \widehat f}
$$
We use the concepts of covering entropy and bracketing entropy in the proofs. The covering entropy $\log N (\epsilon, \mathcal{F}, \|\cdot\|)$ is the logarithm of the minimal number of   $\|\cdot\|$-balls of radius $\epsilon$ needed to cover the set of functions $\mathcal{F}$. The bracketing entropy $\log N_{[ ]} (\epsilon, \mathcal{F}, \|\cdot\|)$ is the logarithm of the minimal number of $\epsilon$-brackets in $ \|\cdot\|$ needed to cover the set of functions $\mathcal{F}$.  An $\epsilon$-bracket $[\ell,u]$ in $ \|\cdot\|$ is the set of functions $f$ with $\ell \leq f \leq u$ and $\|u - \ell \|< \epsilon$.

\section{Proof of Theorems 1 and 2} Throughout this appendix we drop the dependence on $u$ from all the parameters to lighten the notation; for example, $\beta_0$ and $J$ signify $\beta_0(u)$ and $J(u)$.

\subsection{Proof of Theorem 1.}

Step 1. This step shows that $\sqrt{n}(\widehat \beta  -
\beta_0) = O_\Pr(1)$.

By Assumptions   \ref{condition:control} and \ref{condition: selector},  for large
enough  $n$:
$$
1\{S(\widehat \vv)'\widehat \gamma \geq \varsigma\} T \leq
1\{S'\gamma_0\geq\varsigma - \epsilon_n \} T \leq
1\{S'\gamma_0\geq\varsigma/2 \} T \leq 1\{X'\beta_0\geq C + \epsilon'\}T,
$$
$P$-a.e., since
$$
| S(\widehat \vartheta)'\widehat \gamma - S'\gamma_0| T  \leq
\epsilon_n : = L_S ( \|\widehat \vv - \vv_0 \|_{T,\infty} + \|\widehat
\gamma - \gamma_0\|_2) \to_{\Pr} 0,
$$
where $L_S :=(\|\partial_v s \|_{T,\infty} \vee \| s \|_{T,\infty} )$ is a
finite constant by assumption.
Hence, with probability approaching one
$$
\widehat \beta  = \arg \min_{\beta \in \mathbb{R}^{\dim(X)}}
\mathbb{E}_n [\rho_u(Y - X(\widehat \vartheta)'\beta) 1(S(\widehat
\vartheta)'\widehat \gamma \geq \varsigma) T \chi],
$$
where $\chi := 1\{X'\beta_0\geq C + \epsilon'\}$.

Due to convexity of the objective function, it suffices to show that
for any $\epsilon>0$ there exists a finite positive constant
$B_{\epsilon}$ such that
\begin{eqnarray}\label{eq: prob0}
\liminf_{n \to \infty } \Pr \left( \inf_{\|\eta\|_2=1}\sqrt{n} \eta'
\En \Big [\widehat f_{\eta,B_{\epsilon}} \Big ]  > 0 \right) \geq 1-
\epsilon,
\end{eqnarray}
where
$$
\widehat f_{\eta,B_{\epsilon}} (A) := \varphi_u \left\{ Y-X(\widehat
\vartheta)'(\beta_0 + B_\epsilon \eta/\sqrt{n})\right\}
X(\widehat\vartheta) 1 \{S(\widehat \vartheta)' \widehat
\gamma\geq\varsigma \} T \chi.
$$
Let
$$
f(A): =    \varphi_u \left\{ Y-X'\beta_0 \right\} X
1\{S'\gamma_0\geq\varsigma \} T.
$$
Then uniformly in $\|\eta\|_2=1$,
\begin{eqnarray*}
\sqrt{n} \eta' \En[ \widehat f_{\eta,B_{\epsilon}}] & = &  \eta' \Gn[ \widehat f_{\eta,B_{\epsilon}}] + \sqrt{n} \eta' \Ep_P[ \widehat f_{\eta,B_{\epsilon}}] \\
 & =_{(1)} &  \eta'\Gn [f]  + o_\Pr(1) +  \eta' \sqrt{n} \Ep_P[ \widehat f_{\eta,B_{\epsilon}}] \\
 & =_{(2)} &  \eta' \Gn [f]  + o_\Pr(1) +  \eta' J\eta B_{\epsilon} + \eta'\Gn[g] + o_\Pr(1) \\
 & =_{(3)} &  O_\Pr(1) + o_\Pr(1) + \eta' J\eta B_{\epsilon} + O_\Pr(1) + o_\Pr(1),
\end{eqnarray*}
where relations (1) and (2) follow by Lemma \ref{lemma SE} and Lemma
\ref{lemma Expand} with $\widetilde \beta = \beta_0 + B_\epsilon
\eta/\sqrt{n}$, respectively, using that $\|\widehat \vartheta -
\widetilde \vartheta \|_{T,\infty} = o_\Pr(1/\sqrt{n})$, $\widetilde
\vartheta \in \Upsilon$, $\|\widetilde \vartheta -
\vartheta_0\|_{T,\infty} = O_\Pr (1/\sqrt{n})$ and $\| \beta_0 +
B_\epsilon \eta/\sqrt{n} - \beta_0 \|_2 = O(1/\sqrt{n})$;
relation (3) holds by Chebyshev inequality. Since $J$ is positive
definite, with minimal eigenvalue bounded away from zero, the
inequality (\ref{eq: prob0}) follows by choosing $B_{\epsilon}$ as a
sufficiently large constant.

Step 2.  In this step we show the main result.  From the subgradient
characterization of the solution to the quantile regression problem
we have
\begin{equation}
\label{eq foc}
 \sqrt{n} \En\left[ \widehat f \right] = \delta_n;  \ \ \|\delta_n\|_2 \leq  \dim(X) \max_{1 \leq i \leq n} \|T_i X_i\|_2/\sqrt{n} = o_\Pr(1),
\end{equation}
where
$$
\widehat f(A) := \varphi_u \left\{ Y-X(\widehat \vartheta)'\widehat
\beta \right\} X(\widehat\vartheta) 1 \{S(\widehat \vartheta)'
\widehat \gamma\geq\varsigma \} T \chi .
$$

Therefore
\begin{eqnarray*}
o_\Pr(1) = \sqrt{n} \En\left[ \widehat f \right] & = &
 \Gn\left[ \widehat f \right] + \sqrt{n} \Ep_P\left[ \widehat f \right] \\
 & =_{(1)} &  \Gn [f]  + o_\Pr(1) +  \sqrt{n} \Ep_P\left[ \widehat f \right] \\
 & =_{(2)} &  \Gn [f]  + o_\Pr(1) +  J \sqrt{n}(\widehat \beta - \beta_0) + \Gn[g]+
 o_\Pr(1),
\end{eqnarray*}
where relations (1) and (2) follow by Lemma \ref{lemma SE} and Lemma
\ref{lemma Expand} with $\widetilde \beta = \widehat \beta$,
respectively, using that $\|\widehat \vartheta - \widetilde
\vartheta\|_{T,\infty} = o_\Pr(1/\sqrt{n})$, $\widetilde \vartheta \in
\Upsilon$, $\|\widetilde \vartheta - \vartheta\|_{T,\infty} = O_\Pr
(1/\sqrt{n})$ and $\| \widehat \beta - \beta_0\|_2 =
O_\Pr(1/\sqrt{n})$.

Therefore by invertibility of $J$,
$$
 \sqrt{n}(\widehat \beta - \beta_0) = - J^{-1} \Gn(f + g) + o_\Pr(1).
$$
By the Central Limit Theorem, $\Gn(f + g) \to_d  N(0,
\mathrm{Var}_P(f + g))$, so that
$$
 \sqrt{n}(\widehat \beta - \beta_0) \to_d  N(0, J^{-1} \mathrm{Var}_P(f + g)J^{-1}).
$$
 $\square$

\subsection{Proof of Theorem 2}

Step 1. This step shows that $\sqrt{n}(\widehat \beta^e  -
\beta_0) = O_\Pr(1)$ under the unconditional probability $\Pr$.

By Assumptions \ref{condition: selector} and \ref{condition:controlb}, with
probability approaching one,
$$
\widehat \beta^e  = \arg \min_{\beta \in \mathbb{R}^{\dim(X)}}
\mathbb{E}_n [e \rho_u(Y - X(\widehat \vartheta^e)'\beta)
1(S(\widehat \vartheta^e)'\widehat \gamma \geq \varsigma) T \chi ],
$$
where $e$ is the random variable used in the weighted bootstrap
and $\chi = 1(X'\beta_0 \geq C + \epsilon').$
Due to convexity of the
objective function, it suffices to show that for any $\epsilon>0$
there exists a finite positive constant $B_{\epsilon}$ such that
\begin{eqnarray}\label{eq: prob}
\liminf_{n \to \infty } \Pr \left( \inf_{\|\eta\|_2=1}\sqrt{n} \eta'
\En  \Big [\widehat f^e_{\eta,B_{\epsilon}} \Big ]  > 0 \right) \geq
1- \epsilon,
\end{eqnarray}
where
$$
\widehat f^e_{\eta,B_{\epsilon}} (A) := e \cdot \varphi_u \left\{
Y-X(\widehat \vartheta^e)'(\beta_0 + B_\epsilon
\eta/\sqrt{n})\right\} X(\widehat\vartheta^e) 1 \{S(\widehat
\vartheta^e)' \widehat \gamma\geq\varsigma  \}  T \chi .
$$
Let
$$
f^e(A): =   e\cdot \varphi_u \left\{ Y-X'\beta_0 \right\} X 1
\{S'\gamma_0\geq\varsigma  \}T.
$$
Then uniformly in $\|\eta\|_2=1$,
\begin{eqnarray*}
\sqrt{n} \eta' \En[ \widehat f^e_{\eta,B_{\epsilon}}] & = &  \eta' \Gn[ \widehat f^e_{\eta,B_{\epsilon}}] + \sqrt{n} \eta' \Ep_P[ \widehat f^e_{\eta,B_{\epsilon}}] \\
 & =_{(1)} &  \eta'\Gn [f^e]  + o_\Pr(1) +  \eta' \sqrt{n} \Ep_P[ \widehat f^e_{\eta,B_{\epsilon}}] \\
 & =_{(2)} &  \eta' \Gn [f^e]  + o_\Pr(1)+  \eta' J\eta B_{\epsilon} + \eta' \Gn[g^e] + o_\Pr(1) \\
 & =_{(3)} &  O_\Pr(1) + o_\Pr(1) +  \eta' J\eta B_{\epsilon} + O_\Pr(1) + o_\Pr(1),
\end{eqnarray*}
where relations (1) and (2) follow by Lemma \ref{lemma SE} and Lemma
\ref{lemma Expand} with $\widetilde \beta = \beta_0 + B_\epsilon
\eta/\sqrt{n}$, respectively, using that $\|\widehat \vartheta^e -
\widetilde \vartheta^e\|_{T,\infty} = o_\Pr(1/\sqrt{n})$, $\widetilde
\vartheta^e \in \Upsilon$, $\|\widetilde \vartheta^e -
\vartheta_0\|_{T,\infty} = O_\Pr (1/\sqrt{n})$ and $\| \beta_0 +
B_\epsilon \eta/\sqrt{n} - \beta_0 \|_2 = O (1/\sqrt{n})$;
relation (3) holds by Chebyshev inequality.  Since $J$ is
positive definite, with minimal eigenvalue bounded away from zero,
the inequality (\ref{eq: prob}) follows by choosing $B_{\epsilon}$
as a sufficiently large constant.

Step 2.  In this step we show that $\sqrt{n}(\widehat \beta^e -
 \beta_0) = - J^{-1} \Gn( f^e + g^e) + o_\Pr(1)$ under
the unconditional probability $\Pr$.

From the subgradient characterization of the solution to the
quantile regression problem we have \begin{equation}
 \sqrt{n} \En\left[ \widehat f^e \right] = \delta^e_n;  \ \ \|\delta^e_n\|_2 \leq  \dim(X) \max_{1 \leq i \leq n} \|e_i T_i X_i\|_2 / \sqrt{n} = o_\Pr(1),
\end{equation}
where
$$
\widehat f^e(A) := e\cdot \varphi_u \left\{ Y-X(\widehat
\vartheta^e)'\widehat \beta^e \right\} X(\widehat\vartheta^e) 1
\{S(\widehat \vartheta^e)' \widehat \gamma\geq\varsigma  \} T \chi .
$$

Therefore
\begin{eqnarray*}
o_\Pr(1) = \sqrt{n} \En\left[ \widehat f^e \right] & = &
 \Gn\left[ \widehat f^e \right] + \sqrt{n} \Ep_{P}\left[ \widehat f^e \right] \\
 & =_{(1)} &  \Gn [f^e]  + o_\Pr(1) +  \sqrt{n}  \Ep_{P}\left[ \widehat f^e \right] \\
 & =_{(2)} &  \Gn [f^e]  + o_\Pr(1) +  J \sqrt{n}(\widehat \beta^e - \beta_0) + \Gn[g^e]+
 o_\Pr(1),
\end{eqnarray*}
where relations (1) and (2) follow by Lemma \ref{lemma SE} and Lemma
\ref{lemma Expand} with $\widetilde \beta = \widehat \beta^e$,
respectively, using that $\|\widehat \vartheta^e - \widetilde
\vartheta^e\|_{T,\infty} = o_\Pr(1/\sqrt{n})$, $\widetilde \vartheta^e
\in \Upsilon$, $\|\widetilde \vartheta^e - \vartheta_0\|_{T,\infty} =
O_\Pr (1/\sqrt{n})$ and $\| \widehat \beta^e - \beta_0\|_2 =
O_\Pr(1/\sqrt{n})$.

Therefore by invertibility of $J$,
$$
 \sqrt{n}(\widehat \beta^e - \beta_0) = - J^{-1} \Gn(f^e + g^e) + o_\Pr(1).
$$

Step 3. In this final step we establish the behavior of $
\sqrt{n}(\widehat \beta^e - \widehat \beta)$ under $\Pr^e$.
Note that $\Pr^e$ denotes the conditional probability measure,
namely the probability measure induced by draws of $e_1,...,e_n$
conditional on the data $A_1,...,A_n$. By Step 2 of the proof of
Theorem 1 and Step 2 of this proof, we have that under $\Pr$:
$$
 \sqrt{n}(\widehat \beta^e - \beta_0) = -J^{-1} \Gn(f^e + g^e) + o_\Pr(1), \ \sqrt{n}(\widehat \beta - \beta_0) = -J^{-1} \Gn(f + g) + o_\Pr(1).
$$
Hence, under $\Pr$
$$
 \sqrt{n}(\widehat \beta^e - \widehat \beta) =  - J^{-1} \Gn(f^e -f + g^e - g) + r_n = - J^{-1} \Gn( (e-1) (f + g)) + r_n, \ r_n = o_\Pr(1).
$$
Note that it is also true that
$$
r_n = o_{\Pr^e}(1) \text{ in $\Pr$-probability},
$$
where the latter statement means that for every $\epsilon>0$, $
\Pr^e( \|r_n\|_2 > \epsilon) \to_{\Pr} 0. $ Indeed, this follows
from Markov inequality and by
$$
\Ep_{\mathbb{P}}[ \Pr^e( \|r_n\|_2 > \epsilon) ] = \Pr(\|r_n\|_2 >
\epsilon ) = o(1),$$ where the latter holds  by the Law of Iterated
Expectations and $r_n = o_\Pr(1)$.

By the Conditional Multiplier Central Limit Theorem, e.g., Lemma
2.9.5 in van der Vaart and Wellner (1996), we have that  conditional
on the data $A_1,...,A_n$
$$
\Gn( (e-1) (f + g)) \to_d  Z := N( 0, \mathrm{Var}_\Pr(f + g)),
\text{ in $\Pr$-probability},
$$
where the statement means that for each $z \in \Bbb{R}^{\dim(X)}$
$$
\Pr^e( \Gn( (e-1) (f + g))  \leq z) \to_\Pr \mathrm{Pr} ( Z \leq z
).
$$
Conclude that conditional on the data $A_1,...,A_n$
$$
 \sqrt{n}(\widehat \beta^e - \widehat \beta) \to_d N(0, J^{-1} \mathrm{Var}_\Pr(f + g) J^{-1}), \text{ in $\Pr$-probability,}
$$
where the statement means that  for each $z \in \Bbb{R}^{\dim(X)}$
$$
\Pr^e(  \sqrt{n}(\widehat \beta^e - \widehat \beta)  \leq z)
\to_\Pr  \mathrm{Pr} ( - J^{-1} Z \leq z  ).
$$
$\square$

\subsection{Lemma on Stochastic Equicontinuity}

\begin{lemma}[Stochastic equicontinuity]\label{lemma SE} Let  $e \geq 0$ be a positive random variable with $\Ep_P[e] = 1$, $\mathrm{Var}_P[e] = 1,$ and $\Ep_P |e|^{2+\delta} < \infty$ for some $\delta > 0$, that
is independent of $(Y,D,W,Z,X,V)$,  including as a special case
$e=1$, and set, for $A = (e,Y,D,W, Z,X,V)$ and $\chi =
1(X'\beta_0
\geq C + \epsilon')$,
$$
f(A, \vartheta, \beta, \gamma) :=  e \cdot [1(Y \leq
X(\vartheta)'\beta) - u] \cdot X(\vartheta) \cdot 1
(S(\vartheta)'\gamma
\geq \varsigma) \cdot T \cdot \chi.
$$
Under the assumptions of the paper, the following relations are
true.

\begin{itemize}\item[(a)] Consider the set of functions
$$
\mathcal{F} = \{ f(A, \vartheta, \beta, \gamma)'\alpha :
(\vartheta,\beta) \in\Upsilon_0 \times \mathcal{B},  \gamma \in
\Gamma,  \alpha \in \mathbb{R}^{\dim(X)},  \| \alpha\|_2 \leq 1 \},
$$
where $\Gamma$ is an open neighborhood of $\gamma_0$ under the
$\|\cdot\|_2$ metric, $\mathcal{B}$ is an open neighborhood of
$\beta_0$ under the $\|\cdot\|_2$ metric,  $\Upsilon_0$ is the
intersection of $\Upsilon$, defined in Assumption
\ref{condition:control}, with a small neighborhood of $\vv_0$ under
the $\|\cdot\|_{T,\infty}$ metric, which are chosen to be small enough
so that:
$$
 |X(\vartheta)'\beta- X'\beta_0| T  \leq \epsilon'/2, \textrm{
P-a.e.} \ \ \forall (\vartheta, \beta) \in
\Upsilon_0\times\mathcal{B},
$$
where $\epsilon'$ is  defined in Assumptions \ref{condition: selector}. This class is
$P$-Donsker with a square integrable envelope of the form $e$ times
a constant.

\item[(b)] Moreover,  if
$ (\vartheta, \beta, \gamma) \to (\vartheta_0, \beta_0, \gamma_0)
$ in the $\|\cdot \|_{T,\infty} \vee \|\cdot\|_2 \vee \|\cdot\|_2$
metric, then
$$
\| f (A, \vartheta, \beta, \gamma) - f(A, \vartheta_0, \beta_0,
\gamma_0) \|_{P,2}\to 0.
$$

\item[(c)] Hence for any
$ (\widetilde \vartheta, \widetilde \beta, \widehat \gamma)
\to_{\Pr} (\vartheta_0, \beta_0, \gamma_0) $ in the $\|\cdot
\|_{T,\infty} \vee \|\cdot\|_2 \vee \|\cdot\|_2$ metric such that
$\widetilde \vartheta \in \Upsilon_0$ ,
$$
\|\Gn f ( A,\widetilde \vartheta, \widetilde \beta, \widehat \gamma)
- \Gn f (A,\vartheta_0, \beta_0, \gamma_0) \|_2 \to_\Pr 0.
$$

\item[(d)]  For for any
$ (\widehat \vartheta, \widetilde \beta, \widehat \gamma) \to_{\Pr}
(\vartheta_0, \beta_0, \gamma_0) $ in the $\|\cdot \|_{T,\infty}
\vee \|\cdot\|_2 \vee \|\cdot\|_2$ metric, so that
$$
\| \widehat \vartheta - \widetilde \vartheta \|_{T,\infty} =
o_\Pr(1/\sqrt{n}),  \text{ where }  \widetilde \vartheta \in
\Upsilon_0,
$$
we have that
$$
\|\Gn f (A,\widehat \vartheta, \widetilde \beta, \widehat \gamma) -
\Gn f (A,\vartheta_0, \beta_0, \gamma_0)\|_2\to_\Pr 0.
$$
\end{itemize}
\end{lemma}

\textbf{Proof of Lemma \ref{lemma SE}.}   The proof is divided in
proofs of each of the  claims.

%

Proof of Claim (a). The proof proceeds in several steps.

Step 1.  Here we bound the bracketing entropy for
$$
\mathcal{I}_1 = \{ [1(Y \leq X(\vartheta)'\beta) - u] T \chi : \beta
\in \mathcal{B}, \vartheta \in \Upsilon_0 \}.
$$
For this purpose consider a mesh $\{\vv_k\}$  over $\Upsilon_0$ of
$\|\cdot\|_{T,\infty}$ width $\delta$, and a mesh $\{\beta_l\}$ over
$\mathcal{B}$ of $\|\cdot\|_2$ width $\delta$.  A generic bracket
over $\mathcal{I}_1$ takes the form
$$
[i_1^0, i_1^1] =  [  \{1(Y \leq X(\vartheta_k)'\beta_l - \kappa
\delta) - u \} T \chi,   \{1(Y \leq X(\vartheta_k)'\beta_l + \kappa
\delta) - u\} T \chi ],
$$
where $\kappa = L_X \max_{\beta \in \mathcal{B}} \| \beta\|_2 + L_X
$, and $L_X := \| \partial_v x \|_{T,\infty} \vee \| x \|_{T, \infty}.$

Note that this is a valid bracket for all elements of
$\mathcal{I}_1$ induced by any $\vv$ located within $\delta$ from
$\vv_k$ and any $\beta$ located within $\delta$ from $\beta_l$,
since
\begin{eqnarray}
|X(\vv)'\beta - X(\vv_k)'\beta_l|T & \leq &  | (X(\vv) - X(\vv_k))'\beta |
T + | X(\vv_k)'(\beta - \beta_l)| T \nonumber \\
 & \leq & L_X \delta \max_{\beta \in \mathcal{B}} \| \beta\|_2 + L_X \delta \leq \kappa \delta,  \label{eq: converge 1}
\end{eqnarray}
and the $\| \cdot \|_{P,2}$-size of this bracket is given by
\begin{eqnarray*}
\| i_1^0 - i_1^1 \|_{P,2} & \leq & \sqrt{ \Ep_P[ P\{ Y \in[
X(\vartheta_k)'\beta_l \pm  \kappa \delta]  \mid D,W,Z, C, \chi=1\} T ]}  \\
& \leq & \sqrt{ \Ep_P[ \sup_{y \in (C + \kappa\delta, \infty)} P\{ Y
\in[ y \pm  \kappa \delta]  \mid X, C, \chi=1\} T]}
\\
& \leq &
\sqrt{ \| f_{Y}(\cdot \mid \cdot)\|_{T,\infty} 2 \kappa \delta}, \\
  \end{eqnarray*}
provided that $2 \kappa \delta < \epsilon'/2$.  In order to derive
this bound we use the condition $|X(\vartheta)'\beta- X'\beta_0| T
\leq \epsilon'/2, \textrm{ $P$-a.e.} \ \ \forall (\vartheta, \beta)
\in \Upsilon_0\times\mathcal{B}$, so that conditional on $\chi=1$ we
have that $X(\vartheta)'\beta \geq C + \epsilon'/2$; and
\begin{equation*}
P\{ Y \in \cdot  \mid D,W,Z, C,
\chi=1\} = P\{ Y \in \cdot  \mid D,W,Z,V, C, \chi=1\} \\ =  P\{ Y \in
\cdot \mid X, C, \chi=1\} ,
\end{equation*}
 because $V = \vv_0(D,W,Z)$ and the
exclusion restriction for $Z$. Hence, conditional on $X,C$ and
$\chi=1$, $Y$ does not have point mass in the region
$[X(\vartheta_k)'\beta_l \pm \kappa \delta] \subset (C,\infty)$, and
by assumption the density of $Y$ conditional on $X,C$ is uniformly
bounded over the region $(C,\infty)$.

Hence, counting the number of brackets induced by the mesh created above, we arrive at the following relationship between the bracketing entropy of $\mathcal{I}_1$ and the covering entropies of $\Upsilon_0$ and $\mathcal{B}$,
$$
\log N_{[]}(\epsilon, \mathcal{I}_1,  \|\cdot\|_{P,2}) \lesssim \log
N(\epsilon^2, \Upsilon_0, \|\cdot\|_{T,\infty}) + \log N(\epsilon^2,
\mathcal{B}, \|\cdot\|_2) \lesssim 1/(\epsilon^2 \log^4 \epsilon) +
\log(1/\epsilon),
$$
and so $\mathcal{I}_1$ is P-Donsker with a constant envelope.

Step 2.  Similarly to Step 1, it follows that
$$
\mathcal{I}_2 = \{ X(\vv)'\alpha T : \vv \in \Upsilon_0, \alpha \in
\mathbb{R}^{\dim(X)}, \|\alpha\|_2 \leq 1 \}
$$
also obeys a similar bracketing entropy bound
$$
\log N_{[]}(\epsilon, \|\cdot\|_{P,2}) \lesssim 1/(\epsilon^2
\log^4 \epsilon) + \log(1/\epsilon)
$$
with a generic bracket taking the form $[i_2^0,i_2^1] =
 [\{X(\vartheta_k)'\beta_l  - \kappa \delta\}T,  \{X(\vartheta_k)'\beta_l +
\kappa \delta\}T]$. Hence, this class is also P-Donsker with a constant
envelope.

Step 3.  Here we bound the bracketing entropy for
$$
\mathcal{I}_3 = \{ 1(S(\vv)'\gamma \geq \varsigma) T :  \vv \in
\Upsilon_0, \gamma \in \Gamma \}.
$$
For this purpose consider the mesh $\{\vv_k\}$  over $\Upsilon_0$ of
$\|\cdot\|_{T,\infty}$ width $\delta$, and a mesh $\{\gamma_l\}$ over
$\Gamma$ of $\|\cdot\|_2$ width $\delta$.  A generic bracket over
$\mathcal{I}_3$ takes the form
$$
[i_3^0, i_3^1] = [ 1( S(\vartheta_k)'\gamma_l - \kappa \delta \geq
\varsigma)T, 1( S(\vartheta_k)'\gamma_l + \kappa \delta \geq
\varsigma)T ],
$$
where $\kappa = L_S \max_{\gamma \in \Gamma} \| \gamma\|_2 + L_S$,
and $L_S := \| \partial_v s \|_{T, \infty} \vee \| s \|_{T, \infty}.$

Note that this is a valid bracket for all elements of
$\mathcal{I}_3$ induced by any $\vv$ located within $\delta$ from
$\vv_k$ and any $\gamma$ located within $\delta$ from $\gamma_l$,
since
\begin{eqnarray}
| S(\vv)'\gamma - S(\vv_k)'\gamma_l| T & \leq &  | (S(\vv) - S(\vv_k))'\gamma| T + | S(\vv_k)'(\gamma - \gamma_l) | T  \nonumber \\
 & \leq & L_S \delta \max_{\gamma \in \Gamma} \| \gamma\|_2 + L_S \delta \leq \kappa \delta, \label{eq: converge2}
\end{eqnarray}
and the $\| \cdot \|_{P,2}$-size of this bracket is given by
$$
\| i_3^0 - i_3^1\|_{P,2} \leq \sqrt{  P \{| S(\vv_k)' \gamma_l -
\varsigma| T \leq  2 \kappa \delta \}  } \leq \sqrt{ \bar f_S 2 \kappa
\delta},
$$
where $\bar f_S$ is a constant representing the uniform upper bound
on the density of random variable $S(\vv)'\gamma$, where the
uniformity is over  $\vv \in \Upsilon_0$ and $\gamma \in \Gamma$.

Hence, counting the number of brackets induced by the mesh created above, we arrive at the following relationship between the bracketing entropy of $\mathcal{I}_3$ and the covering entropies of $\Upsilon_0$ and $\Gamma$,
$$
\log N_{[]}(\epsilon, \mathcal{I}_3,  \|\cdot\|_{P,2}) \lesssim \log
N(\epsilon^2, \Upsilon_0, \|\cdot\|_{T, \infty}) + \log N(\epsilon^2,
\Gamma, \|\cdot\|_2) \lesssim 1/(\epsilon^2 \log^4 \epsilon) +
\log(1/\epsilon)
$$
and so $\mathcal{I}_3$ is P-Donsker with a constant envelope. \\


Step 4.  In this step we verify the claim (a). Note that $
\mathcal{F} = e \cdot \mathcal{I}_1 \cdot \mathcal{I}_2 \cdot
\mathcal{I}_3. $ This class has a square-integrable envelope under
P.
The class $\mathcal{F}$ is P-Donsker by the following argument.  Note that the
product $\mathcal{I}_1 \cdot \mathcal{I}_2 \cdot \mathcal{I}_3$ of
uniformly bounded classes is P-Donsker, e.g., by Theorem 2.10.6 of
van der Vaart and Wellner (1996). Under the stated assumption the
final product of the random variable $e$ with the P-Donsker class
remains to be P-Donsker by the Multiplier Donsker Theorem, namely
Theorem 2.9.2 in van der Vaart and Wellner (1996).


Proof of Claim (b). The claim follows by the Dominated Convergence
Theorem, since any $f \in \mathcal{F}$ is dominated by a
square-integrable envelope under $P$, and by the following three
facts:
\begin{enumerate}
\item in view of the relation such as (\ref{eq: converge 1}), $1(Y \leq X(\vv)'\beta)T\chi \to 1(Y \leq X'\beta_0)T\chi$
everywhere,
 except for the set $\{ A \in \mathcal{A}: Y = X'\beta_0 \}$ whose measure under
$P$ is zero by $Y$ having a uniformly bounded density conditional on
$X,C$;
\item in view of the relation such as (\ref{eq: converge 1}), $|X(\vv)'\beta T - X'\beta_0 T| \to 0$ everywhere;
\item in view of the relation such as (\ref{eq: converge2}),  $1(S(\vv)'\gamma \geq \varsigma) T \to 1(S'\gamma_0 \geq \varsigma)T$ everywhere, except for the set $\{ A \in \mathcal{A}:  S'\gamma_0 = \varsigma\}$ whose measure under $P$ is zero by $S'\gamma_0$ having a bounded density.
\end{enumerate}

Proof of Claim (c). This claim follows from the asymptotic
equicontinuity of the empirical process $(\Gn [f], f \in
\mathcal{F})$ under the $L_2(P)$ metric, and hence also with respect
to the $\|\cdot\|_{T,\infty} \vee \|\cdot\|_2 \vee \|\cdot\|_2$ metric
in view of Claim (b).

Proof of Claim (d). It is convenient to set $ \widehat f := f (A,
\widehat \vartheta, \widetilde \beta, \widehat \gamma)$ and
$\widetilde f := f (A, \widetilde \vartheta, \widetilde \beta,
\widehat \gamma).$ Note that
\begin{eqnarray*}
| \Gn [\widehat f - \widetilde f]| & \leq &    |\sqrt{n} \En
[\widehat f -
\widetilde f ]| +  |\sqrt{n} \Ep_P (\widehat f - \widetilde f)| \\
& \lesssim &  \sqrt{n} \En[ \widehat \zeta \ ] +  \sqrt{n} \Ep_P [\widehat \zeta \ ]  \\&\lesssim&   \Gn [\widehat \zeta \ ] + 2 \sqrt{n} \Ep_P [\widehat
\zeta \ ],
\end{eqnarray*}
where $|f|$ denote an application of absolute value to each element
of the vector $f$, and  $\widehat \zeta$ is defined by the following
relationship, which holds with with probability approaching one,
\begin{equation} \label{eq: bound diff}
|\widehat f - \widetilde f |  \lesssim  |e| \cdot \| X(\widehat
\vv)  - X(\widetilde \vv)  \|_2 \cdot T +  \widehat g + \widehat h \lesssim
\widehat \zeta := e \cdot L_X \Delta_n + \widehat g + \widehat h, \
\ \Delta_n \geq\| \widehat \vv - \widetilde \vv\|_{T,\infty},
\end{equation}
where $L_X = \| \partial_v x \|_{T,\infty} \vee \| x \|_{T,\infty}$,
and, for some constant $k$,
$$
\widehat g := e \cdot 1\{| Y - X(\widetilde \vv)'\widetilde \beta|
\leq k \Delta_n \} T \chi, \ \ \text{ and } \ \ \widehat h := e\cdot
1\{| S(\widetilde \vv)'\widehat \gamma - \varsigma| \leq k
\Delta_n\} T, \ \
$$
and  $\Delta_n = o (1/\sqrt{n})$ is a deterministic sequence.

 Hence it suffices
to show that the result follows from
\begin{eqnarray}\label{eq: 2 facts}
\Gn  [ \widehat \zeta \ ]= o_\Pr(1),
\end{eqnarray}
and
\begin{eqnarray}\label{eq: 2 facts 2}
 \sqrt{n}
\Ep_P [ \widehat \zeta \ ] = o_\Pr(1).
\end{eqnarray}

Note that since $\Delta_n \to 0$, with probability approaching one,
$\widehat g$ and $\widehat h$ are elements of the function classes
\begin{eqnarray*}
\mathcal{G} &=& \{ e \cdot 1(| Y - X(\vv)' \beta | \leq  k)T \chi :
\vv
\in \Upsilon_0, \beta \in \mathcal{B}, k \in [0,\epsilon'/4] \}, \\
\mathcal{H} &=& \{ e \cdot 1(| S(\vv)' \gamma -\varsigma | \leq  k) T
: \vv \in \Upsilon_0, \gamma \in \Gamma, k \in [0,1]\}.
\end{eqnarray*}
By the argument similar to that in the proof of claim (a), we have
that 
$$
\log N_{[]}(\epsilon, \mathcal{G}, L_2(P)) \lesssim 1/(\epsilon^2
\log^4 \epsilon) \text{ and }  \log N_{[]}(\epsilon, \mathcal{H},
L_2(P)) \lesssim 1/(\epsilon^2 \log^4 \epsilon).
$$
Hence these classes are P-Donsker with unit envelopes. 
%
Let $g = e \cdot 1\{| Y - X(\vv)' \beta|
\leq k \Delta_n \} T \chi,$  and $h = e\cdot
1\{| S(\vv)'\gamma - \varsigma| \leq k
\Delta_n\} T$. Note also that if
$ (\vartheta, \beta, \gamma) \to (\vartheta_0, \beta_0, \gamma_0)
$ in the $\|\cdot \|_{T,\infty} \vee \|\cdot\|_2 \vee \|\cdot\|_2$
metric, then
{\small \begin{eqnarray} \label{eq: small g}
&& \|g\|_{P,2}  \leq \sqrt{ \Ep_P[e^2] \cdot P \{ |Y - X(\vartheta)'\beta |T \leq k \Delta_n\}} \leq \sqrt{ 4 \|f_{Y}(\cdot \mid \cdot)\|_{T,\infty}  k \Delta_n} = o(1), \\
&& \|  h\|_{P,2} \leq \sqrt{ \Ep_P[e^2] \cdot P \{
|S(\vartheta)'\gamma - \varsigma|T \leq k
\Delta_n \}} \leq \sqrt{ 4 \bar  f_S k \Delta_n} = o(1),\label{eq:
small h}
\end{eqnarray}
by the assumption on bounded densities and $\Ep_P[e^2] = 2$.}

%
Conclude that the relation (\ref{eq: 2 facts}) holds by (\ref{eq:
bound diff}), (\ref{eq: small g}), (\ref{eq: small h}), the
P-Donskerity of the empirical processes $( \Gn [h], h \in
\mathcal{H})$ and $(\Gn [g], g \in \mathcal{G})$ and hence their
asymptotic equicontinuity under the $\|\cdot\|_{P,2}$ metric. Indeed,  if
$ (\vartheta, \beta, \gamma) \to (\vartheta_0, \beta_0, \gamma_0)
$ in the $\|\cdot \|_{T,\infty} \vee \|\cdot\|_2 \vee \|\cdot\|_2$
metric,
$$
\| e \cdot L_X \Delta_n +  g + h \|_{P,2} = o(1) \Rightarrow \Gn [\widehat \zeta \ ]
= o_\Pr(1).
$$

%
To show (\ref{eq: 2 facts 2}) note that if
$ (\vartheta, \beta, \gamma) \to (\vartheta_0, \beta_0, \gamma_0)
$ in the $\|\cdot \|_{T,\infty} \vee \|\cdot\|_2 \vee \|\cdot\|_2$
metric,
$$
\| e \cdot L_X \Delta_n +  g + h \|_{P,1}   \leq  \Ep_P|e| \cdot L_X  \Delta_n +
\| g\|_{P,1} + \| h\|_{P,1} = o(1/\sqrt{n}) \Rightarrow   \Ep_P
|\widehat \zeta|  = o_\Pr(1/\sqrt{n}),
$$
since $\Delta_n = o(1/\sqrt{n})$, and 
\begin{eqnarray*} && \| g\|_{P,1}  \leq \Ep_P|e| \cdot P \{ |Y -
X(\vartheta)' \beta | T \leq k \Delta_n \} \leq
 2k \|f_{Y}(\cdot \mid \cdot)\|_{T,\infty} \Delta_n = o(1/\sqrt{n}) \\
&& \| h\|_{P,1} \leq  \Ep_P|e| \cdot P \{ |S(
\vartheta)' \gamma - \varsigma| T \leq k \Delta_n \} \leq 2k
\bar f_S  \Delta_n = o(1/\sqrt{n}),
\end{eqnarray*}
by the assumption on bounded densities.

\vfill

\subsection{Lemma on Local Expansion}

\begin{lemma}[Local expansion] \label{lemma Expand}  Under the assumptions stated in the paper, for
 \begin{eqnarray*}
&& \widehat \delta = \sqrt{n}(\widetilde \beta - \beta_0) =
O_{\Pr}(1); \widehat \gamma = \gamma_0 + o_{\Pr}(1);
\\ && \widehat \Delta(d,w,z) = \sqrt{n}(\widehat \vv(d,w,z)-
 \vv_0(d,w,z)) =  \sqrt{n} \ \En[\ell(A, d,w,z)] + o_{\Pr}(1)  \textrm{ in } \ell^{\infty}(\overline{\mathcal{DR}}),  \\
&&  \|\sqrt{n} \ \En[\ell(A,\cdot)]\|_{T, \infty} =O_{\Pr}(1),
\end{eqnarray*}
we have that
\begin{eqnarray}
& & \sqrt{n} \ \Ep_P \varphi_u\{Y - X(\widehat \vv)'\widetilde \beta \}
X(\widehat \vv) 1 \{S(\widehat\vv)'\gamma \geq \varsigma\} T \chi
\nonumber = J \widehat \delta  +  \sqrt{n} \ \En \left[ g(A)
\right]
 + o_{\Pr}(1),
\end{eqnarray}
where
$$
g(A) =  \int  B(a) \ell(A, d,r) d P(a,d,r), \ \ B (A) := f_{Y}
(X'\beta_0 | X,C) X \dot X'\beta_0 1 (S'\gamma_0 \geq
\varsigma) T.
$$
\end{lemma}

\textbf{Proof of Lemma \ref{lemma Expand}.} With
probability approaching  one,
$$
1\{S(\widehat \vv)'\widehat \gamma \geq \varsigma\} T \leq
1\{S'\gamma_0\geq\varsigma - \epsilon_n \} T  \leq
1\{S'\gamma_0\geq\varsigma/2 \} T \leq 1\{X'\beta_0\geq C +
\epsilon'\} T,
$$
$P$-a.e., by Assumptions \ref{condition:control} and  \ref{condition: selector} since
$$
| S(\widehat \vartheta)'\widehat \gamma - S'\gamma_0| T \leq
\epsilon_n : = L_S ( \|\widehat \vv - \vv_0\|_{T, \infty} + \|\widehat
\gamma - \gamma_0\|_2) \to_{\Pr} 0,
$$
where $L_S =(\|\partial_v s\|_{T, \infty} \vee \|s\|_{T, \infty} )$ is a
finite constant by assumption.

Hence uniformly in $X$ over  $\{ X'\beta_0\geq C + \epsilon'\}$,
\begin{eqnarray*}
& & \sqrt{n} \Ep_P[ \varphi_u\{ Y - X(\widehat \vv)'\widetilde \beta\}
\mid D,W,Z,C] T \nonumber
\\
& &  = f_{Y}(X(\bar{\vv}_X)'\bar \beta_X \mid D,W,Z,C) \{
X(\bar{\vv}_X)'\widehat \delta + \dot{X}(\bar{\vv}_X)'
\bar \beta_X \widehat \Delta(D,W,Z) \} T \nonumber \\
& &  =     f_{Y} (X'\beta_0 \mid D,W,Z,C)\{ X'\widehat \delta +
\dot
X'\beta_0 \widehat \Delta (D,W,Z)\} T + R_X,  \\
& &  =     f_{Y} (X'\beta_0 \mid X,C)\{ X'\widehat \delta + \dot
X'\beta_0 \widehat \Delta (D,W,Z)\} T + R_X,  \\
& &  \bar R = \sup_{\{X : X'\beta_0\geq C + \epsilon'\}} |R_X| =
o_{\Pr}(1),\label{eq: expand 2}
\end{eqnarray*}
where $\bar{\vv}_X$ is on the line connecting $\vartheta_0$ and
$\widehat \vartheta$ and $\bar \beta_X$ is on the line connecting
$\beta_0$ and $\widetilde \beta$.  The first equality follows by
the mean value expansion.   The second equality follows by the
uniform continuity assumption of $f_{Y}(\cdot \mid X,C)$ uniformly
in $X,C$, uniform continuity of $X(\cdot)$ and $\dot{X}(\cdot)$, and
by $\|\widehat \vv - \vv_0\|_{T, \infty} \to_\Pr 0$ and $\|\widetilde
\beta - \beta_0\|_2 \to_\Pr 0$. The third equality follows by
$$f_Y(\cdot \mid D,W,Z,C) = f_Y(\cdot \mid D,W,Z,V,C) = f_Y(\cdot
\mid X,C)$$ because $V = \vv_0(D,W,Z)$ and the exclusion restriction
for $Z$.

Since $f_{Y}(\cdot \mid \cdot)$ and the entries of $X$ and $\dot X$
are bounded, $\widehat \delta = O_\Pr(1),$ and $\|\widehat
\Delta\|_{T, \infty} = O_\Pr(1)$,
 with probability approaching one
\begin{eqnarray}\label{eq: exp1}
 & & \Ep_P [\varphi_u\{Y - X(\widehat \vv)'\widetilde \beta \} X(\widehat \vv) 1 (S(\widehat \vv)'\widehat\gamma \geq \varsigma) T \chi] \notag \\
& & = \Ep_P[f_{Y} (X'\beta_0 | X,C) X X' 1 \{S(\widehat \vv)'\widehat\gamma \geq \varsigma\} T ] \widehat \delta \notag \\
& & +  \Ep_P[f_{Y} (X'\beta_0 | X,C) X \dot X'\beta_0  1 \{
S(\widehat \vv)'\widehat \gamma \geq \varsigma\}   T \widehat \Delta
(D,W,Z) ]+ O_{\Pr}(\bar R).
\end{eqnarray}
Furthermore since
$$
\Ep_P |1 ( S'\gamma_0 \geq \varsigma) - 1 (S(\widehat
\vv)'\widehat\gamma \geq \varsigma)| T
 \leq \Ep_P [ 1 ( S'\gamma_0 \in [\varsigma \pm \epsilon_n] ) T ]  \lesssim \bar f_S \epsilon_n \to_{\Pr}
 0,
$$
where $\bar f_S$ is a constant representing the uniform upper bound
on the density of random variable $S'\gamma_0$,
the expression
(\ref{eq: exp1}) is equal to
\begin{equation*}
J \widehat \delta  +  \Ep_P[f_{Y} (X'\beta_0 | X,C) X \dot
X'\beta_0  1 ( S'\gamma_0 \geq \varsigma) T  \widehat
\Delta(D,W,Z) ]+ O_{\Pr} (\bar f_S \epsilon_n + \bar R).
\end{equation*}
Substituting in $\widehat \Delta(d,w,z) = \sqrt{n} \ \En[ \ell(A,
d,w,z)] + o_{\Pr}(1)$ and interchanging $\Ep_P$ and $\En$, we obtain
\begin{equation*}
\Ep_P[f_{Y} (X'\beta_0 | X,C) X \dot X'\beta_0  1 ( S'\gamma_0
\geq \varsigma)  T \widehat \Delta(D,W,Z) ]  =  \sqrt{n} \ \En[ g(A)]
+ o_{\Pr}(1).
\end{equation*}
The claim of the lemma follows. $\square$

\section{Proof of Theorem 3}

To show claim (1), we first note that by Chernozhukov,
Fern\'andez-Val and Melly (2013),
 \begin{eqnarray*}
&&  \sqrt{n}(\widehat \pi(v) - \pi_0(v)) =   \frac{1}{\sqrt{n}}
\sum_{i=1}^n e_i \Ep_P \left[ f_{D}(R'\pi_0(v) \mid R) R
R'\right]^{-1}   [ v- 1\{D \leq R'\pi_0(v)\}] R + o_P(1),
 \end{eqnarray*}
uniformly over $v \in \T.$
By the Hadarmard differentiability of rearrangement-related
operators in Chernozhukov, Fern\'andez-Val and Galichon (2010), the
mapping $\pi \mapsto \phi_\pi$ from $\ell^{\infty}(\T)^{\dim(R)}$
to $\ell^{\infty}(\overline{\mathcal{DR}})$ defined by $$\phi_\pi(d,r) =
\tau + \int_{\T} 1\{ r' \pi(v) \leq d\} d v$$ is Hadamard differentiable
at $\pi=\pi_0$, tangentially to the set of continuous directions,
with the derivative given by
$$
\dot \phi_{\pi_0}[h]= - f_{D}(d \mid r)  r' h(\vv_0(d,r)),
$$
where $\vv_0(d,r) = \phi_{\pi_0}(d,r)$. Therefore by the Functional
Delta Method (Theorem 20.8 in van der Vaart, 1998), we have that in
$\ell^{\infty}(\overline{\mathcal{DR}})$, for $\widehat \vv(d,r) =
\phi_{\widehat \pi}(d,r)$,
$$
\begin{array}{lll}
 \sqrt{n}(\widehat \vv(d,r) - \vv_0(d,r)) & = & - f_{D}(d \mid r)  r'
\frac{1}{\sqrt{n}} \sum_{i=1}^n e_i \Ep_P \left[
f_{D}(R'\pi_0(\vv_0(d,r)) \mid R) R R'\right]^{-1} \times \\
&  &\times  [ \vv_0(d,r)- 1\{D \leq R'\pi_0(\vv_0(d,r))\}] R +
o_{\Pr}(1). \end{array}
$$
The claim (1) then follows immediately. Also for future reference,
note that the result also implies that
\begin{equation}\label{eq: Gaussian QR}
\sqrt{n}(\widehat \pi (\cdot) - \pi_0(\cdot)) \Rightarrow Z_\pi \
\text{ in } \ell^{\infty}(\T),  \ \ \text{ and } \ \
r'\sqrt{n}(\widehat \pi (\cdot) - \pi_0(\cdot)) \Rightarrow r'Z_\pi
\ \ \text{ in } \ell^{\infty}(\T \times \overline{\R}),
\end{equation}
where $Z_\pi$ is a Gaussian process with continuous sample paths.

 The proof of claim (2) is divided in several steps:

Step 1. In this step we construct $\Upsilon$ and bound  its covering entropy.
Let $C^2_M(\T)$ denote the class of functions $f :\T \to
\Bbb{R}$ with all derivatives up to order 2 bounded by a constant
$M$, including the zero order derivative. The covering entropy of
this class is known to obey $\log N(\epsilon, C^2_M(\T),
\|\cdot\|_{\infty}) \lesssim \epsilon^{-1/2}$. Hence also $ \log
N(\epsilon, \times_{j=1}^{\dim(R)} C^2_M(\T), \|\cdot\|_{\infty})
\lesssim \epsilon^{-1/2}. $ Next construct the set of functions $\Upsilon$ for  some small $m>0$ as:
$$
 \left\{ \tau + T \int_{\T} 1\{ R'\pi(v) \leq D \} dv : \pi =
(\pi_1,...,\pi_{\dim(R)}) \in \times_{j=1}^{\dim(R)}C^2_M(\T),
R'\partial \pi(v) > \frac{m}{1-2\tau}  \text{ $P$-a.e.}  \right\}.
$$
Then for any $\pi$ and $\bar \pi$ obeying the conditions in the
display and such that $\|\pi - \bar \pi\|_{\infty} \leq \delta$,
\begin{eqnarray*}
 & &  T \left|  \int_{\T} 1\{R'\pi(v) \leq D \} dv  - \int_{\T} 1\{R'\bar \pi(v) \leq D \} dv \right |   \\
&&    \leq  T \int_{\T}  1\Big \{R'\pi(v) -D \in [-\|R\|_2 \delta,
\|R\|_2 \delta] \Big \} dv
 \lesssim \frac{1}{m} \|T R\|_2 \delta \lesssim \delta,
\end{eqnarray*}
$P$-a.e., since the density of $r'\pi(V)$, $V \sim U(\T)$, is
bounded above by $1/m$. We conclude that
$$
\log N(\epsilon, \Upsilon, \|\cdot\|_{T,\infty}) \lesssim \log
N(\epsilon, \times_{j=1}^{\dim(R)} C^2_M(\T), \|\cdot\|_{\infty})
\lesssim \epsilon^{-1/2}.
$$

Step 2. In this step we show that there exists $\widetilde \vv \in
\Upsilon$ such that $\|\widetilde \vv - \widehat \vv\|_{T,\infty}=
o_{\Pr} (1/\sqrt{n})$.

We first construct  $\widetilde \pi$ such that
\begin{equation}\label{eq: equivalence}
\sqrt{n} \| \widetilde \pi  - \widehat \pi \|_{\infty} =
o_\mathbb{P}(1), \ \ \text{ and } \ \ \max_{r \in \overline{\mathcal{R}}}
\sqrt{n} \| r' (\widetilde \pi  - \widehat \pi)\|_{\infty} =
o_\mathbb{P}(1), \ \
\end{equation}
where with probability approaching one, $\widetilde \pi \in
\times_{j=1}^{\dim(R)}C^2_M(\T)$ and $R'\partial \widetilde \pi(v)
> m/(1-2\tau)$ $P$-a.e., for some $M$ and some $m>0$.

We construct $\widetilde \pi$ by smoothing $\widehat \pi$ component
by component. Let the components of $\widehat \pi$ be indexed by $1
\leq j \leq \dim(R)$. Before smoothing, we need to extend $\widehat
\pi_j$ outside $\T$. Start by extending the estimand $\pi_{0j}$
outside $\T$ onto the $\epsilon$-expansion $\T^{\epsilon}$
smoothly so that the extended function is in the class $C^3$.  This
is possible by first extending $\partial^3 \pi_{0j}$  smoothly and
then integrating up to obtain lower order derivatives and the
function. Then we extend the estimator $\widehat \pi_j $ to the
outer region by setting $ \widehat \pi_j (v)= \pi_{0j}(v)  + \widehat \pi_j(\tau) - \pi_{0j}(\tau)$ if $v \leq \tau$ and $ \widehat \pi_j (v)= \pi_{0j}(v)  + \widehat \pi_j(1- \tau) - \pi_{0j}(1- \tau)$ if $v \geq 1- \tau$. The extension does not produce a feasible estimator, but it is a useful theoretical device. Note that the extended empirical process
$\sqrt{n} (\widehat \pi_j(v) - \pi_{0j}(v))$ remains to be
stochastically equicontinuous by construction. Then we define $\widetilde \pi_j$ as the
smoothed version of $\widehat \pi_j$, namely $$ \widetilde \pi_j(v)
= \int_{\T^{\epsilon}} \widehat  \pi_j(z) [K((z-v)/h)/h] dz, \ \
v \in \T, $$ where $0 \leq h \leq \epsilon$ is bandwidth such
that $\sqrt{n} h^3 \to 0$ and $\sqrt{n} h^2 \to \infty$; $K: \Bbb{R}
\to \Bbb{R}$ is a third order kernel with the properties:
$\partial^\mu K$ are continuous on $[-1,1]$  and vanish outside of
$[-1,1]$ for $\mu=0,1,2$, $\int K(z) dz =1$, and $\int z^\mu K(z)
dz=0$ for $\mu=1,2$. Such kernel exists and can be obtained by
reproducing kernel Hilbert space methods or via twicing kernel
transformations (Berlinet, 1993, and Newey, Hsieh, and Robins,
2004). We then have \begin{eqnarray*}
\sqrt{n}(\widetilde \pi_j(v) - \widehat \pi_j(v)) && = \int_{\T^{\epsilon}} \sqrt{n} [  \widehat \pi_j(z) - \pi_{0j}(z) - (\widehat \pi_j(v) - \pi_{0j}(v) )] [K([z-v]/h)/h]  dz\\
&& \ \  +  \int_{\T^{\epsilon}} \sqrt{n} ( \pi_{0j}(z) -
\pi_{0j}(v) ) [K([z-v]/h)/h] d z.
\end{eqnarray*}
The first term is bounded uniformly in $v \in \T$ by
$\omega(2h)\|K\|_{\infty} \to_{\Pr} 0$ where
$$
\omega (2h) = \sup_{|z- u| \leq 2h} | \sqrt{n} [  \widehat \pi_j(z)
- \pi_{0j}(z) - (\widehat \pi_j(v) - \pi_{0j}(v) )]| \to_{\Pr} 0,
$$
by the stochastic equicontinuity of the process
$\sqrt{n} (\widehat \pi_j(\cdot) - \pi_{0j}(\cdot))$ over $\T^{\epsilon}$.
The second term is bounded  uniformly in $v \in \T$, up to a
constant, by
$$
\sqrt{n} \| \partial^3 \pi_{0j}\|_{\infty} h^3 \int \lambda^3
K(\lambda) d\lambda \lesssim \sqrt{n} h^3 \to 0.
$$
This establishes the equivalence (\ref{eq: equivalence}), in view of
compactness of $\overline{\R}$.

Next  we show that $\|\partial^2 \widetilde \pi_j\|_{\infty} \leq 2
\|\partial^2 \pi_{0j}\|_{\infty} =: M$ with probability approaching
1. Note that
$$
\partial^2\widetilde \pi_j (v)  - \partial^2 \pi_{0j}(v) = \int_{\T^{\epsilon}}  \widehat \pi_j(z)  [\partial^2K( [z-v]/h)/h^3] dz - \partial^2 \pi_{0j}(v),
$$
which can be decomposed into two pieces:
$$
\begin{array}{l}
n^{-1/2} h^{-2} \int_{\T^{\epsilon}} n^{1/2}(\widehat \pi_j(z)
- \pi_{0j}(z)) [\partial^2K( [z-v]/h)/h] dz\\
 + \int_{{\T^{\epsilon}}}  [ \partial^2 K( [z-v]/h)/h^3] \pi_{0j}(z) dz - \partial^2 \pi_{0j}(v). \end{array}
$$
The first piece is bounded uniformly in $v \in \T$  by $ n^{-1/2}
h^{-2}  \omega(2h) \| \partial^2 K\|_{\infty} \to_\Pr 0, $ while, using
the integration by parts, the second piece is equal to
$$
\int_{{\T^{\epsilon}}} [\partial^2 \pi_{0j}(z) - \partial^2
\pi_{0j}(v) ] [K([z-v]/h)/h] dz.
$$
This expression is bounded in absolute value by
$$
\|K\|_{\infty} \sup_{|z-v|\leq 2 h}  | \partial^2 \pi_{0j}(z) -
\partial^2 \pi_{0j}(v) |\to 0,
$$
by continuity of $ \partial^2 \pi_{0j}$ and compactness of
$\T^{\epsilon}$. Thus, we conclude that  $\|\partial^2\widetilde
\pi_j  -
\partial^2 \pi_{0j}\|_{\infty} \to_{\Pr} 0$, and we can also
deduce similarly that  $\|\partial \widetilde \pi_j  - \partial
\pi_{0j}\|_{\infty} \to_{\Pr} 0$, all uniformly in $1 \leq j \leq
\dim(R)$, since $\dim(R)$ is finite and fixed.

Finally, since by Assumption \ref{Ass: QR}(b)  the conditional
density is uniformly bounded above by a constant, this implies that
$R'\partial \pi_0(v)> k$ $P$-a.e., for some constant $k>0$, and
therefore we also have that with probability approaching one,
$R'\partial \widetilde \pi(v)
>  m/(1-2\tau)$ $P$-a.e. for $m := k (1-2\tau)/2 > 0.$

Next we construct
$$
\widetilde \vv(d,r) = \phi_{\widetilde \pi}(d,r) = \tau +  \int_{\T} 1\{
r'\widetilde \pi(v) \leq d\} d v,
$$
if $(d,r) \in \overline{\D\R}$, and $\widetilde \vv(d,r) = \tau$ otherwise.
Note that by construction  $\widetilde \vv \in \Upsilon$ for some
$M$  with probability approaching one.   It remains to show the
first order equivalence with $\widehat \vv$.

By  the Hadarmard differentiability for the mapping $\phi_\pi$
stated earlier and  by the functional delta method (Theorem 20.8 in
van der Vaart, 1998), $\widetilde \vv$ and $\widehat \vv$ have the
same first order representation in $\ell^{\infty}(\overline{\mathcal{DR}})$,
$$
\sqrt{n}(\widetilde \vv (\cdot) - \vv_0 (\cdot)) =\sqrt{n}(\widehat
\vv (\cdot) - \vv_0 (\cdot))  + o_\mathbb{P}(1),
$$
i.e., $\sqrt{n}\| \widetilde \vv - \widehat \vv \|_{T,\infty}
\to_\mathbb{P} 0$. $\square$

\section{Proof of Theorem 4}

 Claim (1) follows from the results of Chernozhukov, Fern\'andez-Val and Melly (2013). Also for future reference, note
that these results also imply that
\begin{equation}\label{eq: Gaussian DR}
\sqrt{n}(\widehat \pi (\cdot) - \pi_0(\cdot)) \Rightarrow Z_\pi
\text{ in } \ell^{\infty}(\overline{\D}), \ \text{ and } \ \
r'\sqrt{n}(\widehat \pi (\cdot) - \pi_0(\cdot)) \Rightarrow r'Z_\pi
\ \text{ in } \ell^{\infty}(\overline{\mathcal{D}\mathcal{R}}),
\end{equation}
where $Z_\pi$ is a Gaussian process with continuous sample paths.

 The proof of claim (2) is divided in several steps:

Step 1.  In this step we construct $\Upsilon$ and bound  its covering
entropy.  Let $C^2_M(\overline{\D})$ denote the class of functions $f
:\overline{\D} \to \Bbb{R}$ with and all the derivatives up to order 2
bounded by a constant $M$, including the zero order derivative. The
covering entropy of this class is known to obey $\log (\epsilon,
C^2_M(\overline{\D}), \|\cdot\|_{\infty}) \lesssim \epsilon^{-1/2}$.
Hence
$$
\log (\epsilon, \times_{j=1}^{\dim(R)} C^2_M(\overline{\D}),
\|\cdot\|_{\infty}) \lesssim \epsilon^{-1/2}.
$$
Next construct
$$
\Upsilon = \left\{  T \Lambda( R'\pi(D) )  : \pi =
(\pi_1,...,\pi_{\dim(R)}) \in
\times_{j=1}^{\dim(R)}C^2_M(\overline{\D}) \right\}.
$$
Then, for any $\pi$ and $\bar \pi$ obeying the condition in the
definition of the preceding class such that $\|\pi - \bar
\pi\|_{\infty} \leq \delta$,
\begin{eqnarray*}
 & & T \left|\Lambda (R'\pi(D)) -\Lambda (R'\bar \pi(D)) \right| \leq
 \|\partial \Lambda\|_{T,\infty} \sup_{r \in \overline{\R}} \|r\|_{\infty} \delta.
\end{eqnarray*}
 We conclude that
$$
\log N(\epsilon, \Upsilon, \|\cdot\|_{T,\infty}) \lesssim \log
N(\epsilon, \times_{j=1}^{\dim(R)} C^2_M(\overline{\D}),
\|\cdot\|_{\infty}) \lesssim \epsilon^{-1/2}.
$$

Step 2. In this step we show that there exists $\widetilde \vv \in
\Upsilon$ such that $\|\widetilde \vv - \widehat \vv\|_{T,\infty}=
o_{\Pr}(1/\sqrt{n})$.

We first construct  $\widehat \pi$ and $\widetilde \pi$ such that,
\begin{equation}\label{eq: equivalence2}
\sqrt{n} \| \widetilde \pi  - \widehat \pi \|_{\infty} =
o_\mathbb{P}(1), \ \ \text{ and } \ \ \max_{r \in \overline{\R}}
\sqrt{n} \| r' (\widetilde \pi  - \widehat \pi)\|_{\infty} =
o_\mathbb{P}(1), \ \
\end{equation}
where with probability approaching one, $\widetilde \pi \in
\times_{j=1}^{\dim(R)}C^2_M(\overline{\D})$, for some $M$.

We construct $\widetilde \pi$ by smoothing $\widehat \pi$ component
by component. Before smoothing, we extend the estimand $\pi_{0j}$
outside $\overline{\D}$, onto the $\epsilon$-expansion
$\overline{\D}^{\epsilon}$ smoothly so that the extended function is
of class $C^3$. This is possible by first extending the third
derivative of $\pi_{0j}$ smoothly and then integrating up to obtain
lower order derivatives and the function. Then we extend $\widehat
\pi_j $ to the outer region by setting $ \widehat \pi_j (d)=
\pi_{0j}(d) +  \widehat \pi_j (\underline{d}) - \pi_{0j}(\underline{d})$  if $d \leq \underline{d},$ and $ \widehat \pi_j (d)=
\pi_{0j}(d) +  \widehat \pi_j (\overline{d}) - \pi_{0j}(\overline{d})$  if $d \geq \overline{d}$.   The extension does not produce a feasible estimator, but it is a useful theoretical device.  Note that the extended
process $\sqrt{n} (\widehat \pi_j(d) - \pi_{0j}(d))$ remains to be
stochastically equicontinuous by construction. Then we define the smoothed version
of $\widehat \pi_j$ as
$$ \widetilde \pi_j(d) = \int_{\overline{\D}^{\epsilon}} \widehat
\pi_j(z) [K((z-d)/h)/h] dz, \ \ d \in \overline{\D},
$$
where $0 \leq h \leq \epsilon$ is bandwidth such that $\sqrt{n} h^3
\to 0$ and $\sqrt{n} h^2 \to \infty$; $K: \Bbb{R} \to \Bbb{R}$ is a
third order kernel with the properties:  $\partial^\mu K$ are
continuous on $[-1,1]$  and vanish outside of $[-1,1]$ for
$\mu=0,1,2$, $\int K(z) dz =1$, and $\int z^\mu K(z) dz=0$ for
$\mu=1,2$. Such kernel exists and can be obtained by  reproducing
kernel Hilbert space methods or via twicing kernel methods
(Berlinet, 1993, and Newey, Hsieh, and Robins, 2004). We then have
\begin{eqnarray*}
\sqrt{n}(\widetilde \pi_j(d) - \widehat \pi_j(d)) && = \int_{\overline{\D}^{\epsilon}} \sqrt{n} [  \widehat \pi_j(z) - \pi_{0j}(z) - (\widehat \pi_j(d) - \pi_{0j}(d) )] [K([z-d]/h)/h]  dz\\
&& \ \  +  \int_{\overline{\D}^{\epsilon}} \sqrt{n} ( \pi_{0j}(z) -
\pi_{0j}(d) ) [K([z-d]/h)/h] d z.
\end{eqnarray*}
The first term is bounded uniformly in $d \in \overline{\D}$ by
$\omega(2h)\|K\|_{\infty} \to_{\Pr} 0$ where
$$
\omega (2h) = \sup_{|z- u| \leq 2h} | \sqrt{n} [  \widehat \pi_j(z)
- \pi_{0j}(z) - (\widehat \pi_j(d) - \pi_{0j}(d) )]| \to_{\Pr} 0,
$$
by the stochastic equicontinuity of the process
$\sqrt{n} (\widehat \pi_j(\cdot) - \pi_{0j}(\cdot))$ over $\overline{\D}^{\epsilon}$. The second term is bounded  uniformly in $d \in \overline{\D}$, up to
a constant, by
$$
\sqrt{n} \| \partial^3 \pi_{0j}\|_{\infty} h^3 \int \lambda^3
K(\lambda) d\lambda \lesssim \sqrt{n} h^3 \to 0.
$$
This establishes the equivalence (\ref{eq: equivalence2}), in view
of compactness of $\overline{\R}$. 

Next  we show that $\|\partial^2 \widetilde \pi_j\|_{\infty} \leq 2
\|\partial^2 \pi_{0j}\|_{\infty} :=M$ with probability approaching
1. Note that
$$
\partial^2\widetilde \pi_j (d)  - \partial^2 \pi_{0j}(d) = \int_{\overline{\D}^{\epsilon}}  \widehat \pi_j(z)  [\partial^2 K( [z-d]/h)/h^3] dz - \partial^2 \pi_{0j}(d),
$$
which can be decomposed into two pieces:
\begin{eqnarray*}
& n^{-1/2} h^{-2} \int_{\overline{\D}^{\epsilon}} n^{1/2}(\widehat
\pi_j(z)
- \pi_{0j}(z)) [\partial^2 K( [z-d]/h)/h]  dz \\
& + \int_{\overline{\D}^{\epsilon}}  [ \partial^2 K( [z-d]/h)/h^3]
\pi_{0j}(z) dz   - \partial^2 \pi_{0j}(d).
\end{eqnarray*}
The first piece is bounded uniformly in $d \in \overline{\D}$  by $
n^{-1/2} h^{-2}  \omega(2h) \| \partial^2 K\|_{\infty} \to_\Pr 0,$ 
while, using the integration by parts, the second piece is equal to
$$ \int_{{\overline{\D}^{\epsilon}}}  [\partial^2 \pi_{0j}(z) - \partial^2 \pi_{0j}(d) ] [K([z-d]/h)/h] dz , $$
which converges to zero uniformly in $d  \in \overline{\D}$  by the
uniform continuity of $\partial^2 \pi_{0j}$ on
$\overline{\D}^{\epsilon}$ and by boundedness of the kernel function.
Thus  $\|\partial^2\widetilde \pi_j  -
\partial^2 \pi_{0j}\|_{\infty} \to_{\Pr} 0$, and similarly conclude that
$\|\partial \widetilde \pi_j  - \partial \pi_{0j}\|_{\infty}
\to_{\Pr} 0$, where convergence is uniform in $1 \leq j \leq
\dim(R)$, since $\dim(R)$ is finite and fixed.

We then construct $ \ \widetilde \vv (d,r) =  \Lambda(r'\widetilde
\pi(d))$ if $(d,r) \in \overline{\D\R},$ and $\widetilde \vv (d,r) = 0$ otherwise. Note that by the preceding arguments $\widetilde \vv \in
\Upsilon$ for some $M$  with probability approaching one.  Finally,
the first order equivalence
 $\sqrt{n}\| \widetilde \vv - \widehat \vv
\|_{T,\infty} \to_\mathbb{P} 0$ follows immediately from (\ref{eq:
equivalence2}), boundedness of $\|\partial \Lambda\|_{T,\infty}$ and
compactness of $\overline{\R}$.



\section{Computation Details for First Stage Estimators} \label{app: computation}

For the OLS estimator of the control variable in our CQIV estimator,
we run an OLS first stage and retain the  residuals  as the control variable. For the quantile
estimator of the control variable, we run first stage quantile
regressions at each quantile from .01 to .99 in increments of .01, i.e. we set $\tau = .01$.
Next, for each observation, we compute the fraction of the quantile
estimates for which the predicted value
is less than or equal to the observed value.
We then evaluate the standard normal quantile function at this value
and retain the result as the estimate of the control variable. 

For the distribution regression estimator of the control variable,
we first create a matrix $n * n$ of indicators, where $n$ is the
sample size. For each value of the endogenous variable in the data
set $y_{j}$ in columns, each row $i$ gives if the log-expenditure of
the individual $i$ is less or equal than $y_{j}$ $(1(y_{i} \leq
y_{j}$)). Second, for each column $j$ of the matrix of indicators,
we run a probit regression of the column on the exogenous variables.
Finally, the estimate of the control variable for the observation
$i$ is the quantile function of the standard normal evaluated at the
predicted value for the probability of the observation $i=j$.


%

\setcounter{table}{0} 

\begin{table}[tbp]
\label{table: sensitivity}
\begin{center}
\includegraphics[width=1.1\textwidth, height= 1.1\textheight]{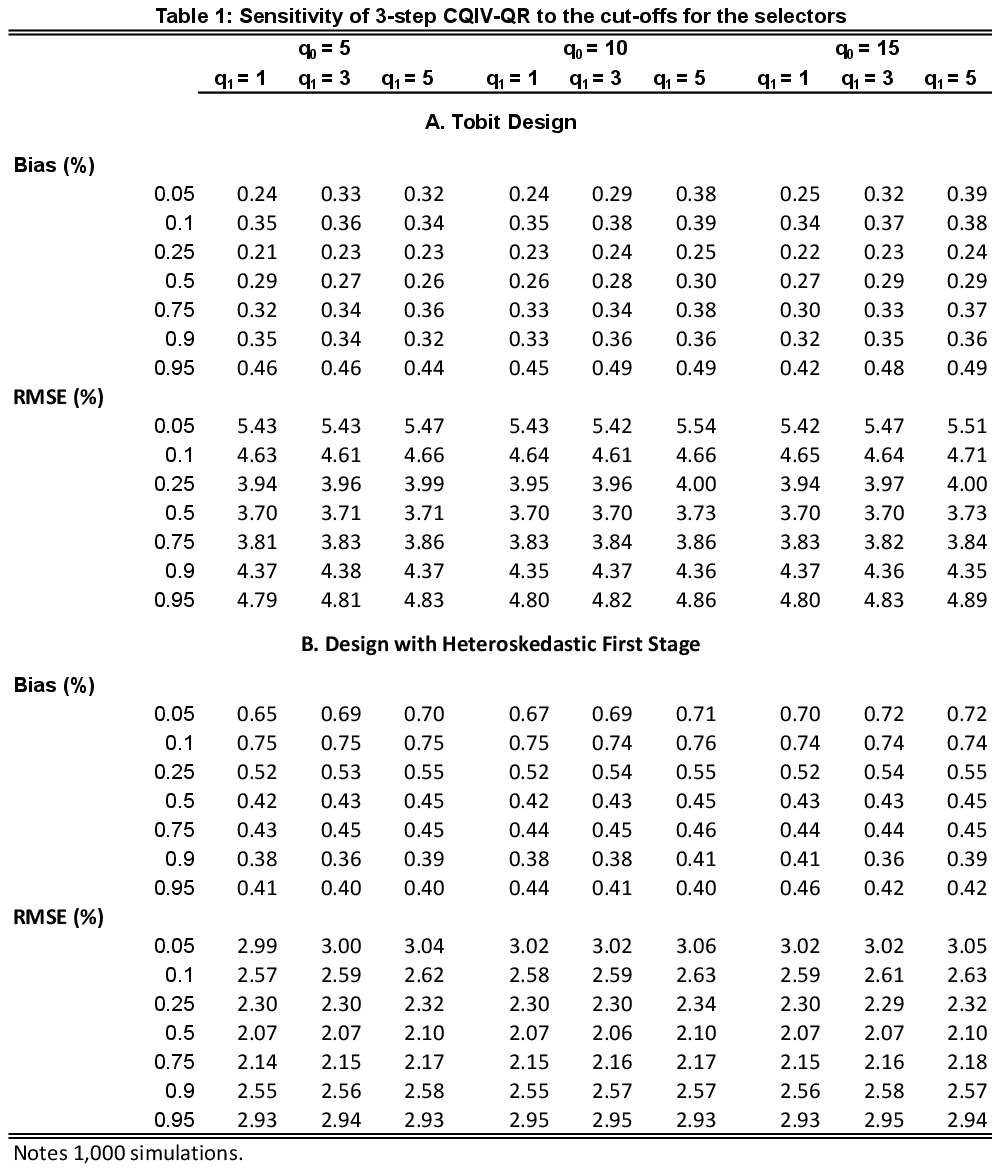}
\end{center}
\end{table}

\begin{table}[tbp]
\label{table: diagnostics}
\begin{center}
\includegraphics[width=1.1\textwidth, height= 1.1\textheight]{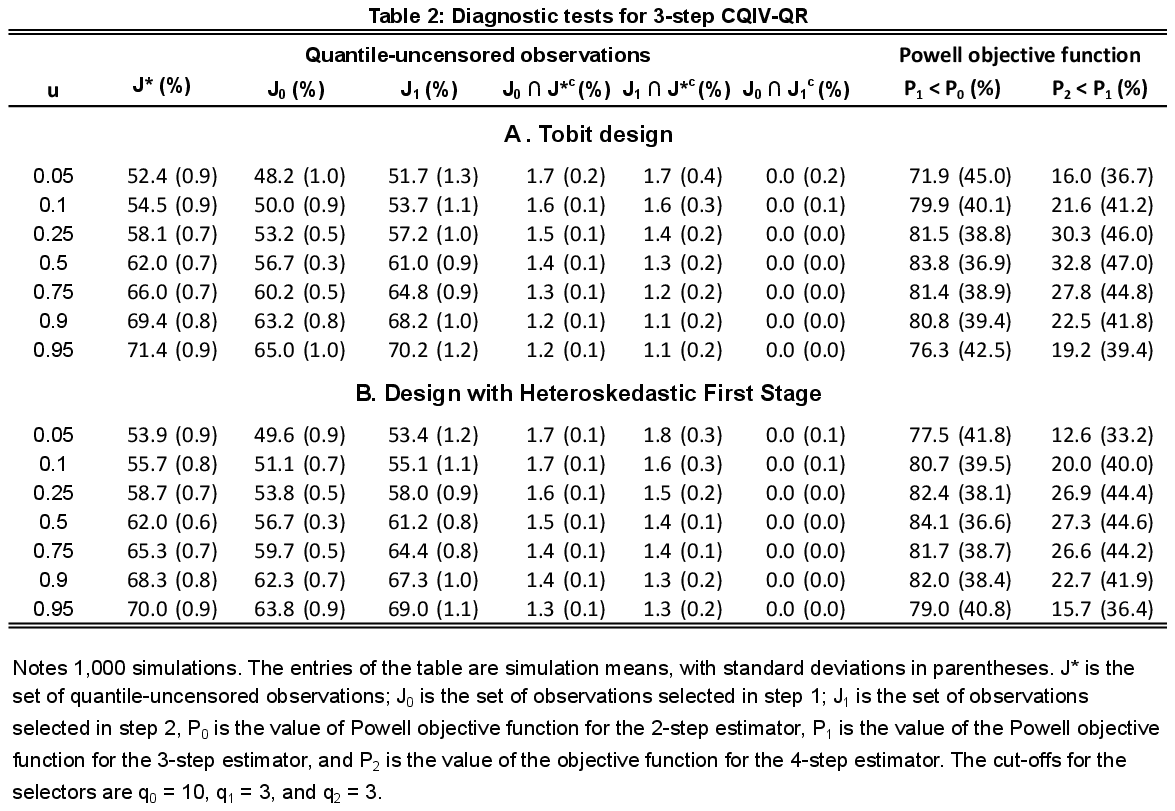}
\end{center}
\end{table}

\end{document}